\documentclass[10pt,a4paper]{article}
\usepackage{mathrsfs}
\usepackage{fancyhdr}
\usepackage[top=30truemm,bottom=30truemm,left=20truemm,right=50truemm]{geometry}
\usepackage{indentfirst}
\usepackage{authblk}
\usepackage{graphicx}% Include figure files
\usepackage{bm,color}% bold math   \textcolor{red} ON
\usepackage[numbers,sort&compress]{natbib}
\usepackage[perpage,para,symbol*]{footmisc}
\usepackage[sort&compress,numbers]{natbib}
\usepackage{doi}%<----------
\usepackage{hyperref}

\renewenvironment{abstract}{\begin{quotation}{\normalsize }}{\end{quotation}}
\makeatletter
\newcommand{\email}[1]{\newcommand{\@email}{E-mail: #1}}
\renewcommand{\maketitle}{
\newpage\null
    \vspace{2em}
 {\LARGE\bfseries\noindent\ignorespaces\@title\par}
 \vspace{1em}%
 {\large\noindent\ignorespaces\@author\par}
 \vspace{2mm}
 {\normalsize\noindent\ignorespaces\@email\par}
\vspace{1em}
}

\renewcommand\@author{\ifx\AB@affillist\AB@empty\AB@author\else
      \ifnum\value{affil}>\value{Maxaffil}\def\rlap##1{##1}%
     \vspace{1mm} \AB@authlist\\\AB@affillist
    \else  \AB@authors\fi\fi}
\makeatother
\usepackage{amsmath,amssymb,amsfonts}
\title{Density constrained TDHF\footnote{To be published as a commemorative ebook focusing on the field of nuclear reaction theory and on the work related to Prof. Joachim Maruhn.}}
\author{V.E. Oberacker and A.S. Umar}
\affil{Department of Physics and Astronomy, Vanderbilt University, Nashville, TN 37235, USA}
\email{volker.e.oberacker@vanderbilt.edu, umar@compsci.cas.vanderbilt.edu}
\begin{document}
% a few definitions
\newcommand{\vcr}{\mbox{${\bf r}\,$}}
\newcommand{\BM}[1]{\mbox{$\displaystyle B^M_{#1}$}}
\newcommand{\bpsi}{\boldsymbol{\psi}}
\newcommand{\bchi}{\boldsymbol{\chi}}

\maketitle

%%%abstract
\begin{abstract}
{\bf Abstract: }
In this manuscript we provide an outline of the numerical methods used in implementing
the density constrained time-dependent Hartree-Fock (DC-TDHF) method and provide a few
examples of its application to nuclear fusion.
In this approach, dynamic microscopic calculations are carried out on a three-dimensional
lattice and there are no
adjustable parameters, the only input is the Skyrme effective NN interaction.
After a review of the DC-TDHF theory and the numerical methods, we present
results for heavy-ion potentials $V(R)$, coordinate-dependent mass parameters $M(R)$, and
precompound excitation energies $E^{*}(R)$ for a variety of
heavy-ion reactions. Using fusion barrier penetrabilities, we calculate
total fusion cross sections $\sigma(E_\mathrm{c.m.})$ for reactions between both stable
and neutron-rich nuclei. We also determine capture cross sections for hot fusion
reactions leading to the formation of superheavy elements.
\end{abstract}
%%%%%%%

%%%keyword
{\bf keywords:} TDHF, DC-TDHF, heavy-ion fusion, neutron-rich nuclei, superheavy elements
%%%%%%

%---------------------------------------------------------------
%---------------------------------------------------------------

\section{Introduction}

The theoretical investigation of heavy-ion interaction potentials $V(R)$ is of fundamental importance
for the study of fusion reactions at energies in the vicinity
of the Coulomb barrier. Radioactive ion beam facilities enable us to study fusion reactions with
exotic neutron-rich nuclei~\cite{balantekin2014}. An important goal of these experiments is to
study the effects of neutron excess ($N-Z$) on fusion. Another theoretical challenge is
to provide guidance to fusion experiments leading to the formation
of new superheavy elements. And finally, fusion of very neutron rich nuclei plays a major role in
nuclear astrophysics where it determines the composition and heating of the crust of
accreting neutron stars.
A recent review of the state of heavy-ion fusion including the DC-TDHF method can be
found in~\cite{back2014}.

In the absence of a practical quantum many-body theory of barrier tunneling, sub-barrier fusion
calculations are commonly performed in two steps; the calculation of a heavy-ion interaction potential
$V(R)$ and a subsequent calculation of tunneling through the barrier.
Theoretical studies of fusion cross sections are often done
by employing phenomenological methods such as the coupled-channels approach~\cite{misicu2006,hagino2012} in which
one uses empirical ion-ion potentials or double-folding potentials calculated with
``frozen density'' approximation and low-lying excitations are incorporated using macroscopic models.
These calculations ignore dynamical effects such as neck formation, particle exchange, pre-equilibrium
GDR and other modes during the nuclear overlap phase of the reaction.
During this phase of the collision the primary underlying mechanism is the dynamical change in the density along the fusion path which modifies the potential energy.
Obviously, this density change is not instantaneous.  For instance, it was shown in Ref.~\cite{simenel2013b} that the
development of a neck due to couplings to  octupole phonons in $^{40}$Ca+$^{40}$Ca could take approximately one
zeptosecond.  As a consequence, the dynamical change of the density is most significant at low energy (near the barrier-top)
where the colliding partners spend enough time in the vicinity of each other with little relative kinetic energy.  At high
energies, however, the nuclei overcome the barrier essentially in their ground-state density.  This energy dependence of
the effect of the couplings on the density evolution was clearly shown in TDHF calculations by Washiyama and Lacroix~\cite{washiyama2008}. This naturally translates into an energy
dependence of the nucleus-nucleus potential~\cite{umar2014a,jiang2014}, similar to what was introduced phenomenologically in the Sao-Paulo
potential~\cite{chamon2002}.  Consequently, the barrier corresponding to near barrier-top energies includes dynamical couplings effects
and can be referred to as a {\it dynamic-adiabatic} barrier, while at high energy the nucleus-nucleus interaction is
determined by a {\it sudden} potential which can be calculated assuming frozen ground-state densities.

We have developed a dynamic microscopic approach, the density constrained time-dependent Hartree-Fock (DC-TDHF)
method~\cite{umar2006b}, to calculate heavy-ion interaction potentials $V(R)$, mass parameters
$M(R)$, and precompound excitation energies $E^{*}(R)$~\cite{umar2009a} directly from microscopic TDHF dynamics.
The idea of constraining the dynamical density during a TDHF collision to find the collective path
corresponding to the dynamical changes of the nuclear density was first introduced in a collaborative
work of Prof.~Maruhn in $1985$~\cite{cusson1985}. This novel approach has facilitated the present
calculations of sub-barrier fusion cross sections,
capture cross sections for superheavy element production, and nuclear astrophysics applications.
The theory has been applied to calculate fusion cross-sections for the fusion of light to heavy systems~\cite{umar2006d,umar2006a,umar2007b,umar2008b,umar2009b,umar2010a, oberacker2010,keser2012,simenel2013a,umar2014a}.
In this paper we will summarize the DC-TDHF method and outline the numerical algorithms involved
in solving the static and dynamic equations of motion. For a variety of heavy-ion reactions,
we discuss numerical calculations of fusion / capture cross-sections and compare these to
experimental data if available.

%---------------------------------------------------------------
%---------------------------------------------------------------

\section{Theoretical formalism}

The theoretical formalism for the
microscopic description of complex many-body quantum systems
and the understanding of the nuclear interactions that result in
self-bound, composite nuclei possessing the observed properties
are the underlying challenges for studying low energy nuclear physics.
The Hartree-Fock approximation
and its time-dependent generalization, the time-dependent Hartree-Fock
theory, has provided a possible means to study the diverse phenomena
observed in low energy nuclear physics
including collective vibrations~\cite{simenel2003,nakatsukasa2005,umar2005a,maruhn2005} and nuclear reactions~\cite{negele1982,simenel2012}.

%---------------------------------------------------------------

\subsection{Time-dependent Hartree-Fock (TDHF) method}

Given a many-body Hamiltonian containing two and three-body
interactions
\begin{equation}
H=\sum_i^N t_i + \sum_{i<j}^Nv_{ij}+\sum_{i<j<k}^Nv_{ijk}\;,
\end{equation}
the time-dependent action $S$ can be constructed as
\begin{equation}
S=\int_{t_1}^{t_2}dt<\Phi(t)|H-i\hbar\partial_t|\Phi(t)>\;.
\end{equation}
Here, $\Phi$ denotes the time-dependent, many-body
wavefunction, $\Phi({\bf r_1,r_2,\ldots,r_{A}};t)$, and $t_i$ is the
one-body kinetic energy operator.
General variation of $S$ recovers the time-dependent Schr\"odinger
equation.
In TDHF approximation the many-body wavefunction is replaced by a single
Slater determinant and this form is preserved at all times.
The determinental form guarantees the antisymmetry required by the Pauli
principle for a system of fermions. In this limit, the
variation of the action yields the most probable time-dependent path
between points $t_1$ and $t_2$ in the multi-dimensional
space-time phase space
\begin{equation}
\delta S=0 \rightarrow \Phi(t)=\Phi_0(t)\;.
\label{variat}
\end{equation}
In practice $\Phi_0(t)$ is chosen to be a Slater determinant
\begin{equation}
\Phi_0(t)=\frac{1}{\sqrt{N!}}det|\phi_{\lambda}(\vcr,t)|\;,
\end{equation}
where $\phi_{\lambda}(\vcr,t)$ are the single-particle states with
quantum numbers $\lambda$.
If the variation in Eq.(\ref{variat}) is performed with respect to
the single-particle states $\phi^{*}_{\lambda}$ we obtain a
set of coupled, nonlinear, self-consistent initial value equations
for the single-particle states
\begin{equation}
h\left( \left\{ \phi_{\mu} \right\} \right) \phi_{\lambda}=i\hbar
\dot{\phi_{\lambda}}
\;\;\;\;\;\;\;\;\;\lambda=1,...,N\;.
\label{tdhf0}
\end{equation}
These are the fully microscopic time-dependent Hartree-Fock equations
which preserve the major conservation laws such as the particle number,
total energy, total angular momentum, etc.
As we see from Eq.(\ref{tdhf0}), each single-particle state evolves in the
mean-field generated by the concerted action of all the other single-particle
states.
Static equations can be obtained from Eq.(\ref{tdhf0}) by taking out a
trivial phase from the single-particle states
\begin{eqnarray}
h(\{\chi_{\mu}\})\chi_{\lambda}&=&\epsilon_{\lambda}\chi_{\lambda}
\nonumber \\
\phi_{\lambda}(\vcr,t)&=&e^{-i\epsilon_{\lambda}t/\hbar}\chi_{\lambda}(\vcr)
\;.
\end{eqnarray}

In TDHF, the initial nuclei are calculated using the static Hartree-Fock (HF) theory. The resulting Slater
determinants for each nucleus comprise the larger Slater determinant describing the colliding
system during the TDHF evolution.
Nuclei are assumed
to move on a pure Coulomb trajectory until the initial separation between the nuclear centers used
in TDHF evolution. Using the Coulomb trajectory we compute the relative kinetic energy at this
separation and the associated translational momenta for each nucleus. The nuclei are then boosted
by multiplying the HF states with
\begin{equation}
\Phi _{j}\rightarrow \exp (\imath\mathbf{k}_{j}\cdot \mathbf{R})\Phi _{j}\;,
\end{equation}
where $\Phi _{j}$ is the HF state for nucleus $j$ and $\mathbf{R}$ is the corresponding
center of mass coordinate
\begin{equation}
\mathbf{R}=\frac{1}{A_{j}}\sum _{i=1}^{A_{j}}\mathbf{r}_{i}\;.
\end{equation}
The Galilean invariance of the TDHF equations assures the evolution of
the system without spreading and the conservation of the total energy for the system.
In TDHF, the many-body state remains a Slater determinant at all times. The final state
is a filled determinant, even in the case of two well separated fragments. This phenomenon
is commonly known as the ``cross-channel coupling'' and indicates that it is not possible
to identify the well separated fragments as distinct nuclei since each single particle state
will have components distributed everywhere in the numerical box~\cite{umar2009c}. In this sense it is
only possible to extract {\it inclusive} (averaged over all states) information from these calculations.

%---------------------------------------------------------------

\subsection{DC-TDHF method}

The concept of using density as a constraint for calculating collective states
from TDHF time-evolution was first introduced in Ref.~\cite{cusson1985}, and used
in calculating collective energy surfaces in connection with nuclear molecular
resonances in Ref.~\cite{umar1985}.

In this approach we assume that a collective state is characterized only by
density  $\rho$, and current $\mathbf{j}$. This state can be constructed
by solving the static Hartree-Fock equations
\begin{equation}
<\Phi_{\rho,\mathbf{j}}|a_h^{\dagger}a_p\hat{H}|\Phi_{\rho,\mathbf{j}}>=0\;,
\end{equation}
subject to constraints on
density and current
\begin{eqnarray*}
<\Phi_{\rho,\mathbf{j}}|\hat{\rho}(\mathbf{r})|\Phi_{\rho,\mathbf{j}}>&=&\rho(\mathbf{r},t) \\
<\Phi_{\rho,\mathbf{j}}|\hat{\jmath}(\mathbf{r})|\Phi_{\rho,\mathbf{j}}>&=&\mathbf{j}(\mathbf{r},t)\;.
\end{eqnarray*}
Choosing $\rho(\mathbf{r},t) $ and $\mathbf{j}(\mathbf{r},t)$ to be the instantaneous TDHF
density and current results in the lowest energy collective state corresponding to the
instantaneous TDHF state $|\Phi(t)>$, with the corresponding energy
\begin{equation}
E_{coll}(\rho(t),\mathbf{j}(t))=<\Phi_{\rho,\mathbf{j}}|\hat{H}|\Phi_{\rho,\mathbf{j}}>\;.
\end{equation}
This collective energy differs from the conserved TDHF energy only by the amount of
internal excitation present in the TDHF state, namely
\begin{equation}
\label{eq:estar}
E^{*}(t)=E_{TDHF} - E_{coll}(t)\;.
\end{equation}
However, in practical calculations the constraint on the current is difficult to implement
but we can define instead a static adiabatic collective state $|\Phi_{\rho}>$ subject to the
constraints
\begin{eqnarray*}
<\Phi_{\rho}|\hat{\rho}(\mathbf{r})|\Phi_{\rho}>&=&\rho(\mathbf{r},t) \\
<\Phi_{\rho}|\hat{\jmath}(\mathbf{r})|\Phi_{\rho}>&=&0\;.
\end{eqnarray*}
In terms of this state one can write the collective energy as
\begin{equation}
\label{eq:4}
E_{coll}=E_{kin}(\rho(t),\mathbf{j}(t))+E_{DC}(\rho(\mathbf{r},t))\;,
\end{equation}
where the density constrained energy $E_{DC}$, and the collective kinetic
energy $E_{kin}$ are defined as
\begin{eqnarray*}
E_{DC}&=&<\Phi_{\rho}|\hat{H}|\Phi_{\rho}> \\
E_{kin}&\approx&\frac{\hbar^2}{2m}\sum_q\int d^{3}r\; \mathbf{j}_q(t)^2/\rho_q(t)\;,
\end{eqnarray*}
where the index $q$ is the isospin index for neutrons and protons ($q=n,p$).

%---------------------------------------------------------------

\subsection{DC-TDHF applications: ion-ion potential, mass parameter,
            and precompound excitation energy}

Recently, we have developed a new method to extract ion-ion interaction potentials directly from
the TDHF time-evolution of the nuclear system~\cite{umar2006b}.
In the DC-TDHF approach
the TDHF time-evolution takes place with no restrictions.
At certain times $t$ or, equivalently, at certain internuclear distances
$\bar{R}(t)$ the instantaneous TDHF density is used to perform a static DFT energy minimization
while constraining the proton and neutron densities to be equal to the instantaneous
TDHF densities. This means we
allow the single-particle wave functions to rearrange themselves in such a way
that the total energy is minimized, subject to the TDHF density constraint.
In a typical DC-TDHF run, we utilize a few
thousand time steps, and the density constraint is applied every $10-20$ time steps.

In essence, DC-TDHF provides us with the
TDHF dynamical path in relation to the multi-dimensional static energy surface
of the combined nuclear system. In this approach
there is no need to introduce constraining operators which assume that the collective
motion is confined to the constrained phase space. In short, we have a self-organizing system which selects
its evolutionary path by itself following the microscopic dynamics.

We refer to the minimized energy as the ``density constrained energy''
$E_{\mathrm{DC}}$.
From Eq.~\ref{eq:4} is is clear that the density constrained energy
plays the role of a collective potential, except for the fact that it contains the binding
energies of the two colliding nuclei. One can thus define the ion-ion
potential as~\cite{umar2006b}
\begin{equation}
V(\bar{R}(t))=E_{\mathrm{DC}}(\rho(\mathbf{r},t))-E_{A_{1}}-E_{A_{2}}\;,
\end{equation}
where  $E_{A_{1}}$ and $E_{A_{2}}$ are the binding energies of two nuclei
obtained from a static Hartree-Fock calculation with the same effective
interaction. For describing a collision of two nuclei one can label the
above potential with ion-ion separation distance $\bar{R}(t)$ obtained during the
TDHF time-evolution. This ion-ion potential $V(\bar{R)}$ is asymptotically correct
since at large initial separations it exactly reproduces $V_{Coulomb}(\bar{R}_{max})$.
All of the dynamical features included in TDHF are naturally included in the DC-TDHF potentials.
These effects include neck formation, particle exchange~\cite{umar2008a,simenel2010,sekizawa2013},
internal excitations, and deformation effects to all order, among others.
Couplings between relative motion and internal structures are known to affect fusion
barriers by dynamically modifying the densities of the colliding nuclei. The effect is expected to be stronger at energies near
the barrier top, where changes in density have longer time to develop than at higher energies.
This gives rise to an energy dependence
of the barriers as predicted by modern time-dependent Hartree-Fock (TDHF) calculations~\cite{washiyama2008,simenel2013b,umar2014a}.

The TDHF evolution also provides us with a coordinate-dependent mass $M(\bar{R})$
which may be obtained using energy conservation for a central collision
\begin{equation}
M(\bar{R})=\frac{2[E_{\mathrm{c.m.}}-V(\bar{R})]}{\dot{\bar{R}}^{2}}\;,
\label{eq:mr}
\end{equation}
where the collective velocity $\dot{\bar{R}}$ is directly obtained from the TDHF evolution.
The $\bar{R}$-dependence of this mass at lower energies is
very similar to the one found in constrained Hartree-Fock calculations~\cite{goeke1983} with a constraint
on the quadrupole moment.
On the other hand, at higher energies the coordinate dependent mass essentially becomes flat,
which is again a sign that most dynamical effects are contained at lower energies.
The peak at small $\bar{R}$ values is
due to the fact that the center-of-mass energy is above the barrier and the
denominator of Eq.~(\ref{eq:mr}) becomes small due to the slowdown of the ions.

The fusion barrier penetrabilities $T_L(E_{\mathrm{c.m.}})$ can be
obtained by numerical integration of the Schr\"odinger equation
for the collective distance coordinate $\bar{R}$, using the heavy-ion potential $V(\bar{R})$
with coordinate dependent mass parameter $M(\bar{R})$.
Alternatively, we can instead use the constant
reduced mass $\mu$ and transfer the coordinate-dependence of the mass to a scaled
potential $V(R)$ using the well known exact point transformation~\cite{goeke1983,umar2009b}
\begin{equation}
dR=\left(\frac{M(\bar{R})}{\mu}\right)^{\frac{1}{2}}d\bar{R}\;.
\label{eq:mrbar}
\end{equation}
The potential $V(R)$, which includes the coordinate-dependent mass effects differs from the
$V(\bar{R})$ only in the interior region of the barrier. Further details can
be found in Ref.~\cite{umar2009b}.

The fusion barrier penetrabilities $T_L(E_{\mathrm{c.m.}})$
are obtained by numerical integration of the Schr\"odinger equation
\begin{equation}
\left[ \frac{-\hbar^2}{2\mu}\frac{d^2}{dR^2}+\frac{L(L+1)\hbar^2}{2\mu R^2}
+V(R)-E_{\mathrm{c.m.}}\right]\psi(R)=0\;,
\label{eq:xfus}
\end{equation}
using the {\it incoming wave boundary condition} (IWBC) method~\cite{rawitscher1964}.
IWBC assumes that once the minimum of the potential is reached fusion will
occur. In practice, the Schr\"odinger equation is integrated from the potential
minimum, $R_\mathrm{min}$, where only an incoming wave is assumed, to a large asymptotic distance,
where it is matched to incoming and outgoing Coulomb wavefunctions. The barrier
penetration factor, $T_L(E_{\mathrm{c.m.}})$, is the ratio of the
incoming flux at $R_\mathrm{min}$ to the incoming Coulomb flux at large distance.
Here, we implement the IWBC method exactly as it is
formulated for the coupled-channel code CCFULL described in Ref.~\cite{hagino1999}.
This gives us a consistent way for calculating cross-sections at energies below and above
the barrier via
\begin{equation}
\sigma_f(E_{\mathrm{c.m.}}) = \frac{\pi}{k^2} \sum_{L=0}^{\infty} (2L+1) T_L(E_{\mathrm{c.m.}})\;.
\label{eq:sigfus}
\end{equation}
At energies above the barrier either the DC-TDHF method or direct TDHF
calculations can be used to determine the fusion cross-sections.
The precompound excitation energy $E^{*}(R(t))$ can be computed from Eq.~\ref{eq:estar}.

The DC-TDHF calculations mentioned above utilize the potential $V(R)$, which includes
the coordinate-dependent mass effects, to find the cross-sections using Eq.~\ref{eq:xfus}.
Recently, a systematic study was
performed to calculate the potential and coordinate dependent mass for non-central TDHF
collisions~\cite{jiang2014}.

%---------------------------------------------------------------

\subsection{Skyrme interaction}

Almost all TDHF calculations have been done using the Skyrme energy density functional.
The Skyrme energy density functional contains terms
which depend on the nuclear density, $\rho$, kinetic-energy density, $\tau$,
spin density, $\mathbf{s}$, spin kinetic energy density, $\mathbf{T}$, and the full spin-current pseudotensor,
$\mathbf{J}$, as
\begin{equation}
E=\int d^{3}r\;{\cal H}(\rho ,\tau,\mathbf{j},\mathbf{{s}},\mathbf{T},\mathbf{J};\mathbf{{r}})\;.
\end{equation}
The time-odd terms ($\mathbf{j}$, $\mathbf{s}$, $\mathbf{T}$) vanish
for static calculations of even-even nuclei, while they are present for odd mass nuclei, in cranking calculations, as well
as in TDHF. The spin-current pseudotensor, $\mathbf{J}$, is time-even and does not vanish for
static calculations of even-even nuclei.
It has been shown~\cite{umar1986a,reinhard1988,umar1989,umar2006c,suckling2010,fracasso2012} that the presence of these
extra terms are necessary for preserving the Galilean
invariance and make an appreciable contribution to the dissipative properties of the collision.
Our TDHF program includes all of the appropriate combinations of time-odd terms in the Skyrme interaction.
In addition, commonly a pairing force is added to incorporate pairing interactions for nuclei. The
implementation of pairing for time-dependent collisions is currently an unresolved problem although
small amplitude implementations exist~\cite{tohyama2002a,tohyama2002b,ebata2010}.
However, for reactions with relatively high excitation this is not expected to be a problem.

%---------------------------------------------------------------
%---------------------------------------------------------------

\section{Numerical methods}

In this section we discuss the numerical details of performing TDHF calculations
of nuclear collisions as well as the density-constraint method, which is crucial
for ion-ion potential calculations. Relatively recently it has become feasible, for the first time,
to perform TDHF calculations on a
3D Cartesian grid without any symmetry restrictions
and with much more accurate numerical methods~\cite{umar2006c,maruhn2014,guo2008}.
At the same time the quality of effective interactions has also been substantially
improved~\cite{chabanat1998a,kluepfel2009,kortelainen2010,kortelainen2012}.

%---------------------------------------------------------------

\subsection{Discrete variation and lattice equations}

The lattice solution of differential equations on a discretized
mesh of independent variables may be viewed to proceed in two
steps: (1) Obtain a discrete representation of the functions and
operators on the lattice. (2) Solve the resulting lattice equations
using iterative techniques.
Step (1) is
an interpolation problem for which we could take advantage of the
techniques developed using the basis-spline functions~\cite{umar1991b,umar1991a}.
The use of the
basis-spline collocation method leads
to a matrix-vector representation on the collocation lattice with
a metric describing the transformation properties of the collocation
lattice.

In order to obtain a
set of lattice equations which preserve the conservation laws
associated with the continuous equations it is essential to develop
a modified variational approach.
This goal is achieved by performing a variation to the
discretized form of a conserved quantity, i.e. total energy.
Consequently, the resulting equations will preserve all of the
conserved quantities on the lattice.
For the TDHF equations we consider
a general discretized form of the action
\begin{equation}
S=\int\, dt \sum_{\alpha\beta\gamma}
\Delta V_{\alpha\beta\gamma}\left\{
{\cal H}(\alpha\beta\gamma)-
\left[i\hbar \sum_{\mu}\psi_{\mu}^*(\alpha\beta\gamma)
\frac{\partial\psi_{\mu}}{\partial t}(\alpha\beta\gamma)\right]
\right\}\;,
\end{equation}
where indices $\alpha,\beta,\gamma$ denote the lattice points in
three-dimensional space, and $\Delta V_{\alpha\beta\gamma}$ is the
corresponding infinitesimal volume element. Due to the presence of
derivative operators in the Hamiltonian the explicit form of these
expressions will depend non-locally on the lattice indices.
The general variation, which preserves the properties of the continuous
variation, is given by
\begin{equation}
\frac{\delta\psi_{\mu}^{*}(\alpha\beta\gamma)}
{\delta\psi_{\lambda}^{*}(\alpha'\beta'\gamma')}=
\frac{1}{\Delta V_{\alpha\beta\gamma}}
\delta_{\lambda\mu}\delta_{\alpha'\alpha}\delta_{\beta'\beta}
\delta_{\gamma'\gamma}\;.
\label{var3}
\end{equation}
Until recently, most HF and TDHF calculations have been performed using
finite-difference lattice techniques. The details of the discrete
variation for the finite-difference case are given in Refs.~\cite{umar1989,koonin1977}.
Below we outline a procedure for using the BSCM for the numerical solution
of HF and TDHF equations. Further details of the BSCM is published
elsewhere~\cite{umar1991b}.

%---------------------------------------------------------------

\subsection{Basis-splines}

\begin{figure}[!hbt]
\begin{center}
\includegraphics[width=10.0cm]{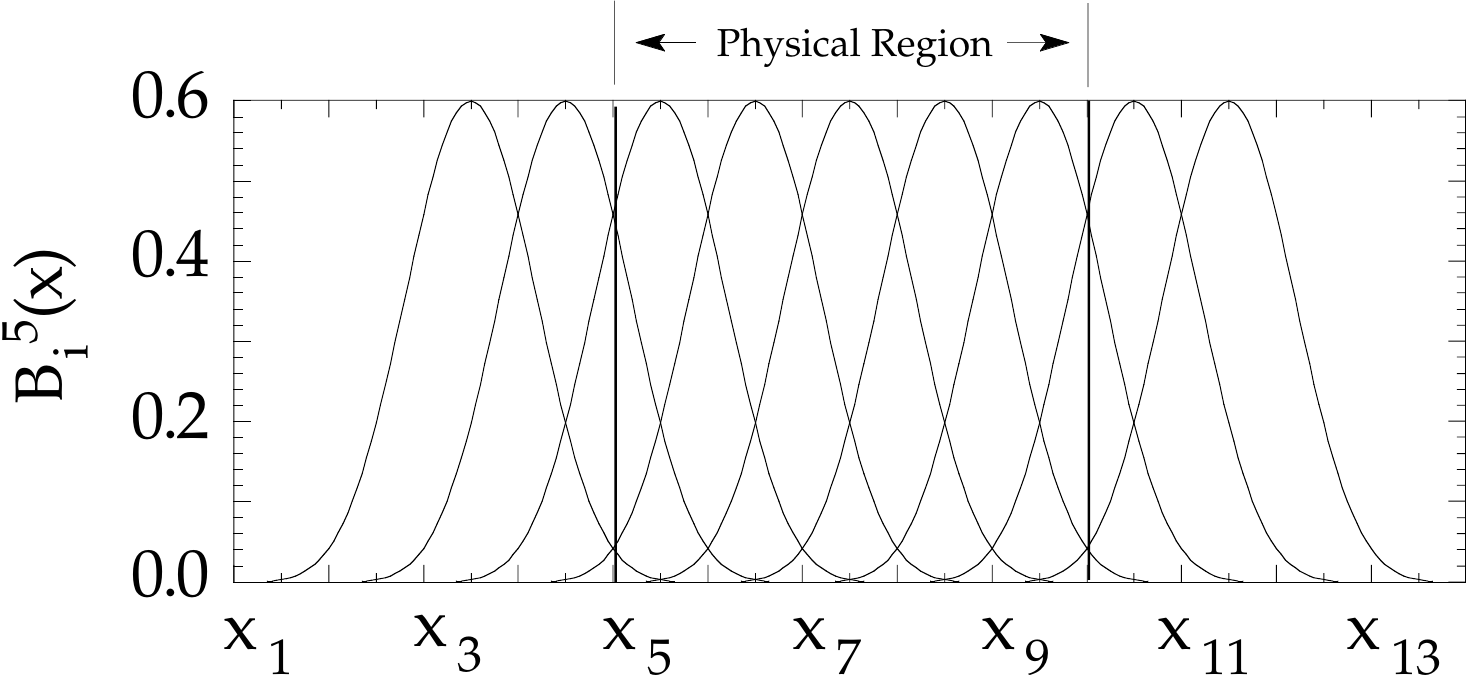}
\end{center}
\caption{\protect\footnotesize A region of space with physical boundaries located at knots $x_M$ and
$x_{M+N}$ for $M=5$ and $N=5$.
The B-spline $B_1^M$ which begins at the first knot $x_1$ has its tail
in the physical region. The last
B-spline which begins within the physical boundaries is $B^M_{N+M-1}$.
It extends up to the last knot $x_{N+2M-1}$.}
\label{fig0}
\end{figure}

Given a set of points or {\it knots} denoted by the set $\{x_i\}$
a basis-spline (B-spline denoted by \BM{i}) function of order $M$ is
constructed from continuous piecewise polynomials of order $M-1$~\cite{boor1978}.
B-splines have continuous
derivatives up to $(M-2)^{nd}$ derivative and a discontinuous $(M-1)^{st}$
derivative.
We only consider odd
order splines or even order polynomials for reasons related to the
choice of the collocation points.
The $i^{th}$ B-spline
is nonzero only in the interval
$(x_i,x_{i+M})$. This property is commonly referred to as limited
support. The knots are the points
where polynomials that make up the B-spline join. In the interval
containing the tail region B-splines fall off very rapidly to zero.
An example of order $M=5$ splines extending over a physical region is illustrated in Fig.\ref{fig0}.
We can
also construct exact derivatives of B-splines provided the
derivative order does not
exceed $M-1$.

A continuous function $f(x)$, defined in the interval
$(x_{min},x_{max})$, can
be expanded in terms of B-spline functions as
\begin{equation}
f(x)=\sum_i\BM{i}(x)c^i\;,
\label{bexp}
\end{equation}
where quantities $c^i$ denote the expansion coefficients.
We can solve for the expansion coefficients in terms of a given
(or to be determined)
set of function values evaluated at a set of data points more commonly
known as {\it collocation points}. There are a number of ways to choose
collocation points \cite{boor1978,umar1991b}, however,
for odd order B-splines a
simple choice is to place one collocation point
at the center of each knot interval
within the physical boundaries
\begin{equation}
x_{\alpha}=\frac{x_{\alpha+M-1}+x_{\alpha+M}}{2}\;\;,
\;\;\alpha=1,\ldots,N.
\end{equation}
Here, $x_M=x_{min}$, $x_{N+M}=x_{max}$, and $N$ is the number of collocation
points.
We can now
write a linear system of equations by evaluating (\ref{bexp}) at these
collocation points
\begin{equation}
f_{\alpha}=\sum_iB_{\alpha i}c^i\;,
\label{nobnd}
\end{equation}
where $f_{\alpha}\equiv f(x_{\alpha})$, and $B_{\alpha i}\equiv \BM{i}
(x_{\alpha})$.
In order to solve for the expansion coefficients the
matrix ${\bf B}$ needs to be inverted.
However, as it stands
matrix ${\bf B}$ is not a square matrix, since the total number of
B-splines with a nonzero extension in the physical region is $N+M-1$.
In order to
perform the inversion we need to introduce additional linear equations
which represent the boundary conditions imposed on $f(x)$ at the
two boundary points, $x_M$ and $x_{M+N}$.
The essence of the lattice method is to eliminate the expansion
coefficients $c^i$ using this inverse matrix. The details of using
the boundary conditions (or periodic boundaries) and inverting the resulting square matrix are
discussed in Ref.~\cite{umar1991b}.
Following the inversion the coefficients are given by
\begin{equation}
c^i=\sum_{\alpha}\left[ {\bf B}^{-1} \right]^{i\alpha}f_{\alpha}\;.
\label{btilm}
\end{equation}
One can
trivially show that all local functions will have a local representation
in the finite dimensional collocation space
\begin{equation}
f(x)\longrightarrow f_{\alpha}\;.
\label{funeq}
\end{equation}

The collocation representation of the operators can be obtained by
considering the action of an operator ${\cal O}$ onto a function $f(x)$
\begin{equation}
{\cal O}f(x)=\sum_i[{\cal O}\BM{i}(x)]c^i\;.
\end{equation}
If we evaluate the above expression at the collocation points $x_{\alpha}$
we can write
\begin{equation}
[{\cal O}f]_{\alpha}=\sum_i[{\cal O}B]_{\alpha i}c^i\;.
\label{nouse}
\end{equation}
Substituting from Eq.~(\ref{btilm}) for the coefficients $c^i$ we obtain
\begin{eqnarray}
[{\cal O}f]_{\alpha}&=&\sum_{i\beta}[{\cal O}B]_{\alpha i}
\left[ {\bf B}^{-1} \right]^{i\beta}f_{\beta} \nonumber \\
&=&\sum_{\beta}O_{\alpha}^{\beta}f_{\beta}\;,
\label{opeq}
\end{eqnarray}
where we have defined the collocation space matrix representation of
the operator ${\cal O}$ by
\begin{equation}
O_{\alpha}^{\beta}=\sum_i[{\cal O}B]_{\alpha i}
\left[{\bf B}^{-1}\right]^{i\beta}\;.
\label{colop}
\end{equation}
Notice that the construction of the collocation space operators can be
performed once and for all at the beginning of a calculation, using only
the given knot sequence and collocation points.
Due to the presence of the inverse in Eq.~(\ref{colop}) the matrix
$O$ is not sparse. In practice, operator ${\cal O}$ is
chosen to be a differential operator such as $d/dx$ or $d^2/dx^2$.
By a similar construction it is also possible to obtain the appropriate
integration weights on the collocation lattice~\cite{umar1991b}.

%---------------------------------------------------------------

\subsection{Discrete HF equations}

Since the detailed derivation
of the BSCM representation of the TDHF equations
involve many terms that are
present in the energy functional, we will only show few terms.
The three-dimensional expansion in terms of B-splines is a simple
generalization of Eq.~(\ref{bexp})
\begin{equation}
\psi_{\lambda}(x,y,z)=\sum_{ijk}c^{ijk}_{\lambda}B_{i}(x)B_{j}(y)B_{k}(z)\;.
\end{equation}
The knots and collocation points for each coordinate can be different.
With the appropriate definition of boundary conditions all of the
discretization techniques discussed in the previous section can be
generalized to the three-dimensional space. The details of this procedure
are given in~\cite{umar1991b}.

As an example for a local term let us consider a part of the $t_0$
contribution to the total energy
\begin{equation}
\frac{t_0}{2}(1+\frac{x_0}{2})\int d^3r\rho^2=\frac{t_0}{2}(1+\frac{x_0}{2})
\sum_{\alpha\beta\gamma}w^{\alpha}w^{\beta}w^{\gamma}
[\rho(\alpha\beta\gamma)]^2\;,
\label{var1}
\end{equation}
where on the right-hand side we have written the discretized form on a
collocation lattice with collocation weights denoted by $w$.
Here, $\alpha,\beta,\gamma$ represent the collocation points in $x,y,$ and
$z$ directions, respectively.
In order to be able to perform the
variation with respect to the single-particle states
$\psi_{\lambda}^{*}$ we rewrite equation (\ref{var1}) explicitly
\begin{equation}
\frac{t_0}{2}(1+\frac{x_0}{2})
\sum_{\alpha\beta\gamma}w^{\alpha}w^{\beta}w^{\gamma}
\sum_{\mu\nu}\psi_{\mu}^{*}\psi_{\mu}\psi_{\nu}^{*}\psi_{\nu}\;.
\label{var2}
\end{equation}
Using Eq.(\ref{var3}) in the variation of Eq.(\ref{var2}) we obtain
(after replacing the primed indices with unprimed ones) the contribution
\begin{equation}
t_0(1+\frac{x_0}{2})\rho(\alpha\beta\gamma)
\psi_{\lambda}(\alpha\beta\gamma)\;,
\end{equation}
where we have rewritten a summation as the total density.
The same procedure can be carried
out for the nonlocal terms in the energy density. A typical term
is illustrated below
\begin{eqnarray*}
({\mbox{\boldmath $\nabla$}}\mathbf\psi_{\lambda}^{\pm})_{\alpha \beta \gamma}&=&
\sum_{\alpha '}D_{\alpha}^{\alpha '}\psi_{\lambda}^{\pm}(\alpha '\beta
\gamma)\,\hat{\imath}
+\sum_{\beta '}D_{\beta}^{\beta '}\psi_{\lambda}^{\pm}(\alpha \beta '
\gamma)\,\hat{\jmath}
\\
&+&\sum_{\gamma '}D_{\gamma}^{\gamma '}\psi_{\lambda}^{\pm}(\alpha \beta
\gamma ')\,\hat{k}
\end{eqnarray*}
where the matrices ${\bf D}$ denote the first derivative matrices in
$x,y,z$ directions (they can be different although the notation does not
make this obvious), calculated as described in the previous section.
Finally, the HF equations can be written as matrix-vector equations on
the collocation lattice
\begin{equation}
h\psi_{\lambda}^{\pm}\longrightarrow
{\bf h}\cdot \mbox{{\boldmath $\psi$}}_{\lambda}^{\pm}\;.
\label{hfeq}
\end{equation}
The essence of this construction is that the terms in the single-particle
Hamiltonian ${\bf h}$ are matrices in one coordinate and diagonal in
others. Therefore, ${\bf h}$ need not be stored as a full matrix, which
allows the handling of very large systems directly in memory. The
details of this procedure are discussed below.

%---------------------------------------------------------------

\subsection{Solution of the discrete equations}

In this subsection we will outline some of the numerical methods developed
for the solution of the
discretized HF and TDHF equations. The subsection is divided
into two parts; the first part discusses the static iteration methods and
the solution of the field equations and in particular the powerful
damped relaxation method \cite{bottcher1989}. In addition we discuss the
implementation of external constraints on the HF equations. The
second part of the subsection introduces a number of time-evolution
methods used in our calculations. Typical numerical accuracies are
also discussed.

\subsubsection{Static solutions}

The solution of the HF equations (\ref{hfeq}) represent the problem
of finding the few lowest eigenvalues of a very large Hamiltonian matrix.
Furthermore, due to the fact that
we are dealing with a self-consistent problem the
matrix elements must be recalculated at every iteration.
However, in
practice the matrix elements need not be stored. Instead, one can make
use of the inherent sparsity to dynamically construct the operation
of the single-particle Hamiltonian onto a statevector. The basic operation
is
\begin{equation}
\bpsi'={\bf h}\cdot\bpsi\;,
\end{equation}
where the construction of the right hand side is done by explicitly
programming the required linear combinations of the elements of $\bpsi$
to give $\bpsi'$. In this approach the only storage requirements are for
the statevectors and small matrices present in the Hamiltonian.

The lattice equations are solved by using the
damped relaxation method described in Refs.~\cite{umar1985,bottcher1989}.
A simple way for introducing the damped relaxation method is by pointing
out its resemblance to the so-called imaginary time method. A more
formal discussion is given in Ref.~\cite{bottcher1989} where the generalization
to the relativistic Dirac equation is also introduced.
We start with the TDHF equations
\begin{equation}
i\hbar\frac{\partial\bpsi_{\lambda}}{\partial t}={\bf h}(t)\bpsi_{\lambda}\;.
\end{equation}
In terms of the discretized time $t_n=n\Delta t$ the solution at time
$n+1$ can be obtained from time $n$ by (see below)
\begin{equation}
\bpsi_{\lambda}^{n+1}=e^{-i\Delta t{\bf h}^n/\hbar}\bpsi_{\lambda}^n\;,
\end{equation}
where ${\bf h}^n$ is the single-particle Hamiltonian at the $n^{th}$
iteration. The imaginary time-step method consists of the transformation
$\Delta t\rightarrow -i\Delta t$
\begin{equation}
\bchi_{\lambda}^{n+1}=e^{-x_0 ({\bf h}^n-\epsilon_{\lambda}^n)}
\bchi_{\lambda}^n\;,
\label{imexp}
\end{equation}
where $x_0=\Delta t/\hbar$, and we have taken out a trivial phase from
$\bpsi_{\lambda}^n$.
The expansion of the exponential to first order in $x_0$ yields the
imaginary time iteration scheme
\begin{equation}
\bchi_{\lambda}^{n+1}={\cal O}{\bf [} \bchi_{\lambda}^n-
x_0({\bf h}^n-\epsilon_{\lambda}^n)\bchi_{\lambda}^n {\bf ]}\;,
\label{imtime}
\end{equation}
where ${\cal O}$ stands for Gram-Schmidt orthonormalization, which is
necessary to ensure the orthonormality of the single-particle states
at each iteration. In Eq.~(\ref{imtime}) the index $n$ is no longer
associated with time and it simply becomes an iteration counter.
It is clear from Eq.~(\ref{imexp}) that the exponential acts as  a filter
in selecting the lowest eigenvalues of ${\bf h}$ and leads to the
minimization of the HF energy.
The generalization of the imaginary time method, where we introduce
the damping matrix ${\bf D}$, results in the damped relaxation method
\begin{equation}
\bchi_{\lambda}^{n+1}={\cal O}{\bf [} \bchi_{\lambda}^n-
x_0 {\bf D}(E_0)({\bf h}^n-\epsilon_{\lambda}^n)\bchi_{\lambda}^n {\bf ]}\;.
\end{equation}
The damping operator ${\bf D}$ is chosen to be
\begin{eqnarray*}
{\bf D}(E_0)&=&\left[ 1+\frac{{\bf T}}{E_0}\right]^{-1} \\
&\approx&
\left[ 1+\frac{{\bf T}_x}{E_0}\right]^{-1}
\left[ 1+\frac{{\bf T}_y}{E_0}\right]^{-1}
\left[ 1+\frac{{\bf T}_z}{E_0}\right]^{-1}\;,
\end{eqnarray*}
where ${\bf T}$ denotes the kinetic energy operator. Limits can be
established for the ranges of the parameters $x_0$ and $E_0$ \cite{bottcher1989},
but in practice fine-tuning is necessary for optimal performance.
Two convergence criteria are used in practical calculations; one being the
fractional change in the HF energy
\begin{equation}
\Delta E^n=\frac{E^{n+1}-E^n}{E^n}\;,
\end{equation}
and other the fluctuations in energy
\begin{equation}
\eta\equiv\sqrt{<H^2>-<H>^2}\;.
\end{equation}
The fluctuations are a more stringent condition than
the simple energy difference between two iterations.
In practice, we have required $\eta$ to be less than $10^{-6}$.
For this value of the energy fluctuation the fractional change
in the HF energy is about $10^{-13}$.

The calculation of the HF Hamiltonian also requires the
evaluation of the direct Coulomb contribution. However, since
the calculation of the three-dimensional Coulomb integral is very costly,
we solve instead the corresponding differential equation
\begin{equation}
\nabla^2U_C(\vcr)=-4\pi e^2\rho_p(\vcr)\;.
\end{equation}
Details of solving the Poisson equation using the
BSCM is given in Ref.~\cite{umar1991b}.

\subsubsection{Constrained HF calculations}

It is sometimes desirable to solve the static HF equations away from the
global minimum in energy. Such situations usually arise in the study of
fission barriers and in the study of long-lived superdeformed states of
nuclei that can be formed during low energy heavy-ion collisions. These
methods have been instrumental for the understanding of
the formation of nuclear molecules~\cite{umar1985}. All of these cases require
the existence of a stable minimum which does not coincide with the
ground state configuration. The usual approach is to
study the HF energy of a nuclear system by keeping certain macroscopic
degrees of freedom at pre-specified values. This results in a
multi-dimensional {\it energy surface} from which extremum values can be
obtained. The reliability of these results depend strongly on the
correct identification of the relevant macroscopic degrees of freedom.
However, as we will see below a special constrained HF method, {\it density
constrained HF}, has also been developed which allows the minimization of
the energy along a TDHF trajectory.

The goal is to devise an iteration scheme such that the expectation value
of an arbitrary operator $\hat{Q}$ does not change from one static
iteration to next
\begin{equation}
\sum_{\lambda}<\chi_{\lambda}^{n+1}|\hat{Q}|\chi_{\lambda}^{n+1}>=
\sum_{\lambda}<\chi_{\lambda}^n|\hat{Q}|\chi_{\lambda}^n>\;.
\end{equation}
Furthermore, we require this expectation value to be a fixed number $Q_0$.
A procedure can be developed by using Lagrange multipliers that are
dynamically adjusted~\cite{umar1985}. We start with the static HF iteration
scheme modified by the addition of a constraint	(we have omitted the
damping matrix ${\bf D}$ in the equations below for simplicity)
\begin{equation}
\bchi_{\lambda}^{n+1}={\cal O}{\bf [} \bchi_{\lambda}^n-
x_0({\bf h}^n+\lambda\hat{Q}-\epsilon_{\lambda}^n)\bchi_{\lambda}^n {\bf ]}\;.
\end{equation}
In Ref.\cite{umar1985} we give a set of exact equations which preserve the
expectation of the constraining operator to order $x_0^2$. However, these
equations involve the calculation of exchange terms and may become costly.
Instead, one can develop a simpler iterative scheme as follows. Perform
an intermediate step
\begin{equation}
\bchi_{\lambda}^{n+1/2}={\cal O}{\bf [} \bchi_{\lambda}^n-
x_0({\bf h}^n+\lambda^n\hat{Q}-\epsilon_{\lambda}^n)
\bchi_{\lambda}^n {\bf ]}\;,
\end{equation}
and calculate the difference
\begin{equation}
\delta Q^{n+1/2}=
\sum_{\lambda}<\chi_{\lambda}^{n+1/2}|\hat{Q}|\chi_{\lambda}^{n+1/2}>-
\sum_{\lambda}<\chi_{\lambda}^n|\hat{Q}|\chi_{\lambda}^n>\;.
\end{equation}
In analogy with the exact case the Lagrange parameter $\lambda$ is
altered to reduce this difference
\begin{equation}
\lambda^{n+1}=\lambda^n+c_0\frac{\delta Q^{n+1/2}}{2x_0\sum_{\lambda}
<\chi_{\lambda}^n|\hat{Q}^2|\chi_{\lambda}^n>+d_0}\;,
\end{equation}
where $c_0$ and $d_0$ are empirical parameters replacing the exchange terms.
In terms of these intermediate states the $(n+1)st$ step is given by
\begin{equation}
\bchi_{\lambda}^{n+1}={\cal O}{\bf [} \bchi_{\lambda}^{n+1/2}-
x_0(\lambda^{n+1}-\lambda^n+\delta\lambda^n)\hat{Q}
\bchi_{\lambda}^{n+1/2} {\bf ]}\;,
\end{equation}
where
\begin{equation}
\delta\lambda^n=\frac{\sum_{\lambda}<\chi_{\lambda}^n|\hat{Q}|\chi_{\lambda}^n>
-Q_0}{2x_0\sum_{\lambda}<\chi_{\lambda}^n|\hat{Q}^2|\chi_{\lambda}^n>+d_0}\;.
\end{equation}

The extension of the method which allows the entire density to be constrained
is straightforward.
In this case we would like to constrain a continuous density
\begin{equation}
\rho^n(\vcr)=\sum_{\lambda}|\chi_{\lambda}^n(\vcr)|^2\;,
\end{equation}
to be equal to $\rho_0(\vcr)$.
The constraining operator $\hat{Q}$ becomes the density operator
$\hat{\rho}(\vcr)$ defined in the single-particle space
\begin{equation}
<\chi_{\lambda}^n|\hat{\rho}(\vcr)|\chi_{\lambda}^n>=|\chi_{\lambda}^n(\vcr)|^2
\end{equation}
and the product $\lambda\hat{Q}$ is replaced by an integral
\begin{equation}
\lambda\hat{Q}\longrightarrow \int d^3r\,\lambda(\vcr)\hat{\rho}(\vcr)=
\lambda(\vcr)\;.
\end{equation}
The last equality is due to the fact that in coordinate space
$\hat{\rho}(\vcr)$ is a delta function.
An iterative scheme for $\lambda^n(\vcr)$ is given by
\begin{equation}
\lambda^{n+1}(\vcr)=\lambda^n(\vcr)+c_0
\frac{\delta\rho^{n+1/2}}{2x_0\rho^n(\vcr)+d_0}\;,
\end{equation}
where
\begin{equation}
\delta\rho^{n+1/2}(\vcr)\equiv \rho^{n+1/2}(\vcr)-\rho_0(\vcr)
\end{equation}
is obtained from half-time iteration step
\begin{equation}
\bchi_{\lambda}^{n+1/2}={\cal O}{\bf [} \bchi_{\lambda}^n-
x_0({\bf h}^n+\lambda^n(\vcr)-\epsilon_{\lambda}^n)
\bchi_{\lambda}^n {\bf ]}\;.
\end{equation}
Note that in these equations we require that the density remain equal
to $\rho_0(\vcr)$ at every iteration and not just at the final step.
Using these wavefunctions the full iteration can be written as
\begin{equation}
\bchi_{\lambda}^{n+1}={\cal O}{\bf [} \bchi_{\lambda}^{n+1/2}-
x_0(\lambda^{n+1}(\vcr)-\lambda^n(\vcr)+\delta\lambda^n(\vcr))
\bchi_{\lambda}^{n+1/2} {\bf ]}\;,
\end{equation}
where
\begin{equation}
\delta\lambda^n(\vcr)=c_0\frac{\rho^n(\vcr)-\rho_0(\vcr)}{2x_0\rho_0(\vcr)+d_0}\;.
\end{equation}

During the density constrained HF iterations the single-particle states
readjust to minimize the energy while the initial density is kept fixed.
In practical calculations the parameter $x_0$ has been replaced by the
damping operator and the constants $c_0$ and $d_0$ were chosen to be $1.9$
and $5\times10^{-5}$, respectively.

\subsubsection{Time evolution}

The formal solution of the TDHF equations (\ref{tdhf0}) is
\begin{equation}
\psi_{\lambda}(t)=U(t,t_0)\psi_{\lambda}(t_0)\;,
\end{equation}
where we have omitted the spatial coordinates for
simplicity and the time propagator $U(t,t_0)$ is given by
\begin{equation}
U(t,t_0)={\cal T}\,exp\left[\frac{-i}{\hbar}\int_{t_0}^t\,dt'h(t')\right]\;.
\end{equation}
The quantity ${\cal T}$ denotes time-ordering which is necessary in the
general case. In practical calculations we discretize time as
\begin{equation}
t_n=n\Delta t\;\;\;\;\;\;\;\;\;\;\;\;\;\;\;n=0,1,2,...N\;,
\end{equation}
and express the evolution operator in successive infinitesimal pieces
\begin{equation}
U(t,t_0)=U(t,t_{N-1})U(t_{N-1},t_{N-2})...U(t_1,t_0)\;.
\end{equation}
In this case the time-ordering operator can be ignored.
For three-dimensional calculations the exponential operator is expanded as
a Taylor series
\begin{equation}
U(t_{n+1},t_n)\approx \left( 1+\sum_{k=1}^{K}\frac{(-i\Delta t\,{\bf
h}/\hbar)^k}{k!}\right) \;. \end{equation}
The expansion of the operator requires repeated applications of ${\bf h}$
onto the wavefunctions. In practice, only $6-8$ terms are needed for
the conservation of the norm at 1 part in $10^{-10}$ level during the
entire time-evolution. The expansion method is also attractive due to the
fact that it only involves matrix vector operations which could be
easily customized for vector or parallel computers.

%---------------------------------------------------------------

\section{Numerical results}

%---------------------------------------------------------------

\subsection{Fusion Calculations for Neutron-Rich and Stable Nuclei}

In fusion experiments with neutron-rich radioactive ion beams, the dynamics
of the neutron skin usually enhances the sub-barrier fusion cross
section over that predicted by a simple static barrier penetration model, but
in some cases suppression of fusion is also observed. Most recently, we have investigated
sub-barrier fusion and pre-equilibrium giant resonance excitation between
various tin + calcium isotopes~\cite{oberacker2012,oberacker2013} and calcium + calcium isotopes~\cite{keser2012}. Finally,
we have studied sub-barrier fusion reactions between both stable and neutron-rich
isotopes of oxygen and carbon~\cite{umar2012a,desouza2013,rudolph2012} that occur in the neutron
star crust. In all cases, we have found
good agreement between the measured fusion cross sections and the DC-TDHF results.
This is rather remarkable given the fact that the only input in DC-TDHF is the
Skyrme effective N-N interaction, and there are no adjustable parameters.

In several experiments, large enhancements of sub-barrier fusion yields have
been observed for systems with positive Q values for neutron transfer.
Recently, a series of experiments has been carried out
with radioactive $^{132}$Sn beams and with stable $^{124}$Sn beams on
$^{40,48}$Ca targets~\cite{kolata2012}. It turns out that the $^{40}$Ca+Sn systems
have many positive Q values for neutron-pickup while all the Q values for
$^{48}$Ca+Sn are negative. However, the data analysis reveals that the fusion
enhancement is not proportional to the magnitudes of those Q values.

\begin{figure}[!htb]
\begin{center}
\includegraphics[width=6.0cm]{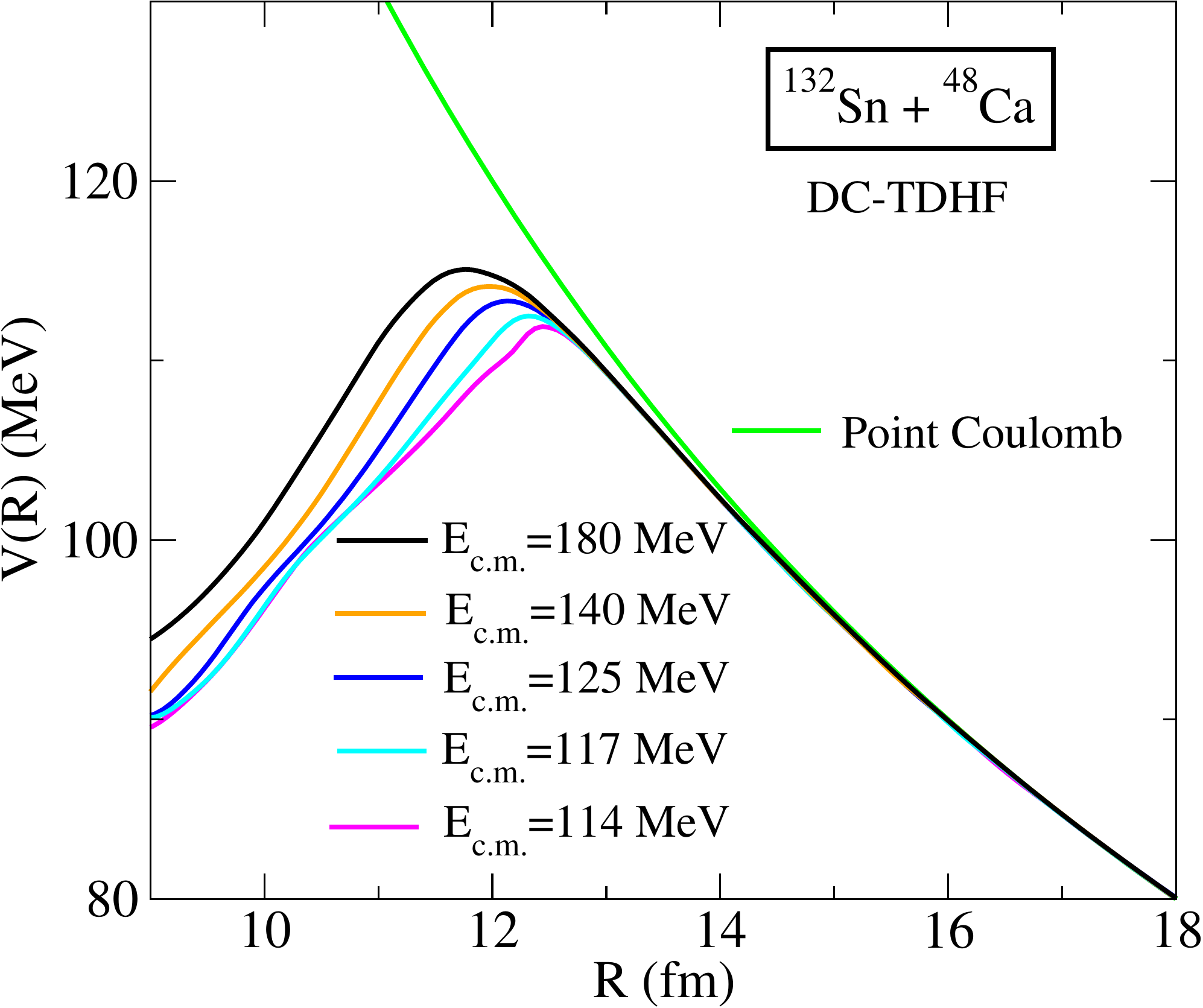}\hspace{0.5cm}
\includegraphics[width=6.0cm]{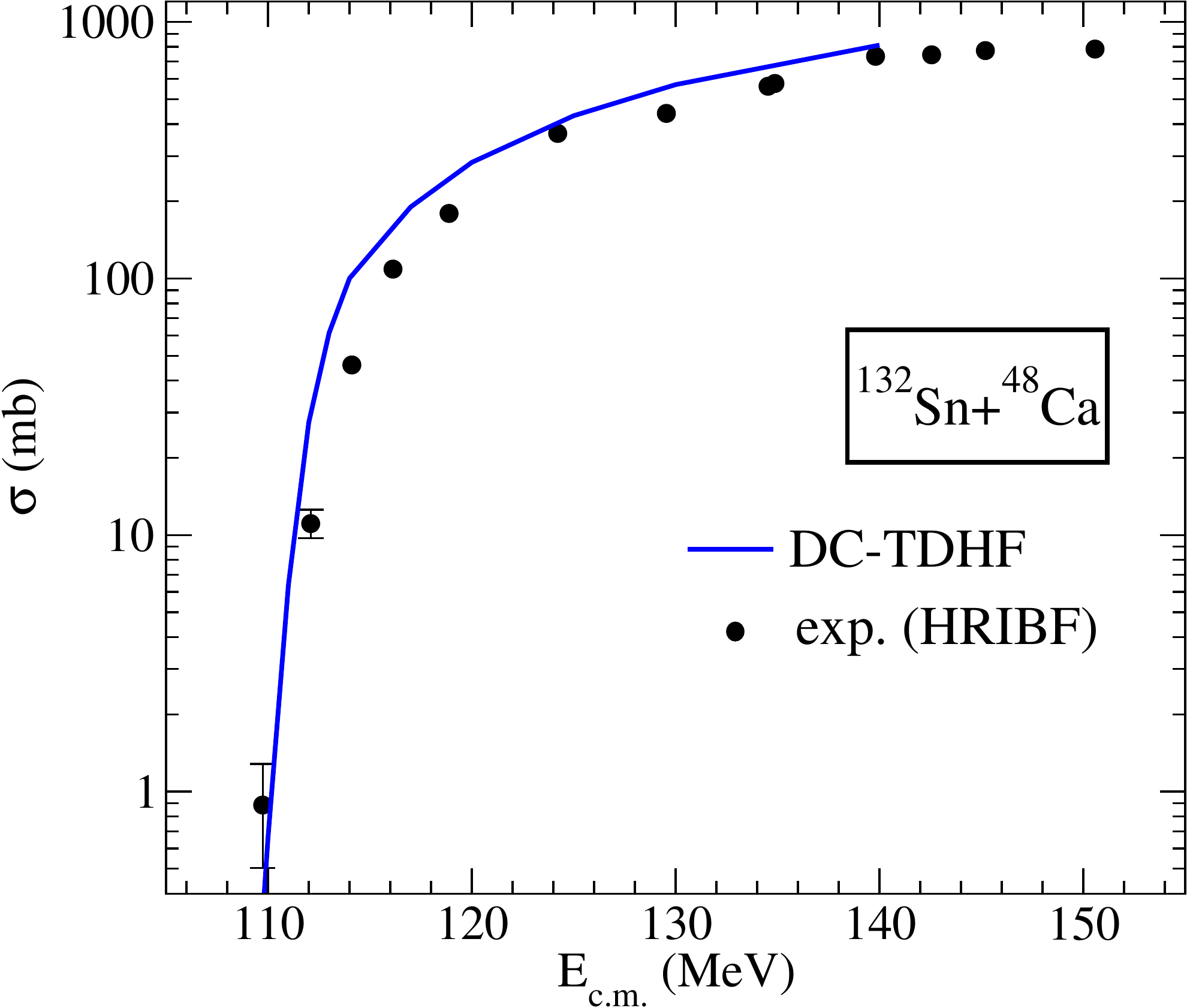}
\end{center}
\caption{\protect\footnotesize Left: DC-TDHF calculation~\cite{oberacker2012,oberacker2013}
of heavy-ion interaction potentials for the neutron-rich system $^{132}$Sn+$^{48}$Ca.
The potentials are shown at five $E_\mathrm{c.m.}$ energies. Right: Total
fusion cross section as a function of center-of-mass energy. The experimental
data are taken from Ref.~\cite{kolata2012}.}
\label{fig:Sn+Ca_1}
\end{figure}

First, we consider the neutron-rich system $^{132}$Sn+$^{48}$Ca. Results for
the heavy-ion interaction potential $V(R)$ with the DC-TDHF method~\cite{oberacker2012,oberacker2013}
are shown in Fig.~\ref{fig:Sn+Ca_1} on the left side.
In general, the ion-ion potentials are energy-dependent, and they are
shown here at five $E_\mathrm{c.m.}$ energies. Our results demonstrate that in these heavy
systems the potential barrier height increases dramatically with increasing
energy, and the barrier peak moves inward towards smaller $R$-values.
On the right side of Fig.~\ref{fig:Sn+Ca_1} we show the calculated fusion cross section
as a function of $E_\mathrm{c.m.}$. In this case, we have used the theoretical
cross sections obtained with the energy-dependent DC-TDHF potentials~\cite{oberacker2012}. We can see that our
theoretical cross sections agree remarkably well with the experimental data.

Particularly puzzling is the experimental observation of a sub-barrier fusion
enhancement in the system $^{132}$Sn+$^{40}$Ca
as compared to more neutron-rich system $^{132}$Sn+$^{48}$Ca.
This is difficult to understand because the $8$ additional neutrons in
$^{48}$Ca should increase the attractive strong nuclear interaction
and thus lower the fusion barrier, resulting in an enhanced sub-barrier
fusion cross section. But the opposite is found experimentally.
A coupled channel analysis~\cite{kolata2012} of the fusion data with
phenomenological heavy-ion potentials yields cross sections that are one order
of magnitude too small at sub-barrier energies, despite the fact that
these ion-ion potentials contain many adjustable parameters.
\begin{figure}[!htb]
\begin{center}
\includegraphics[width=6.0cm]{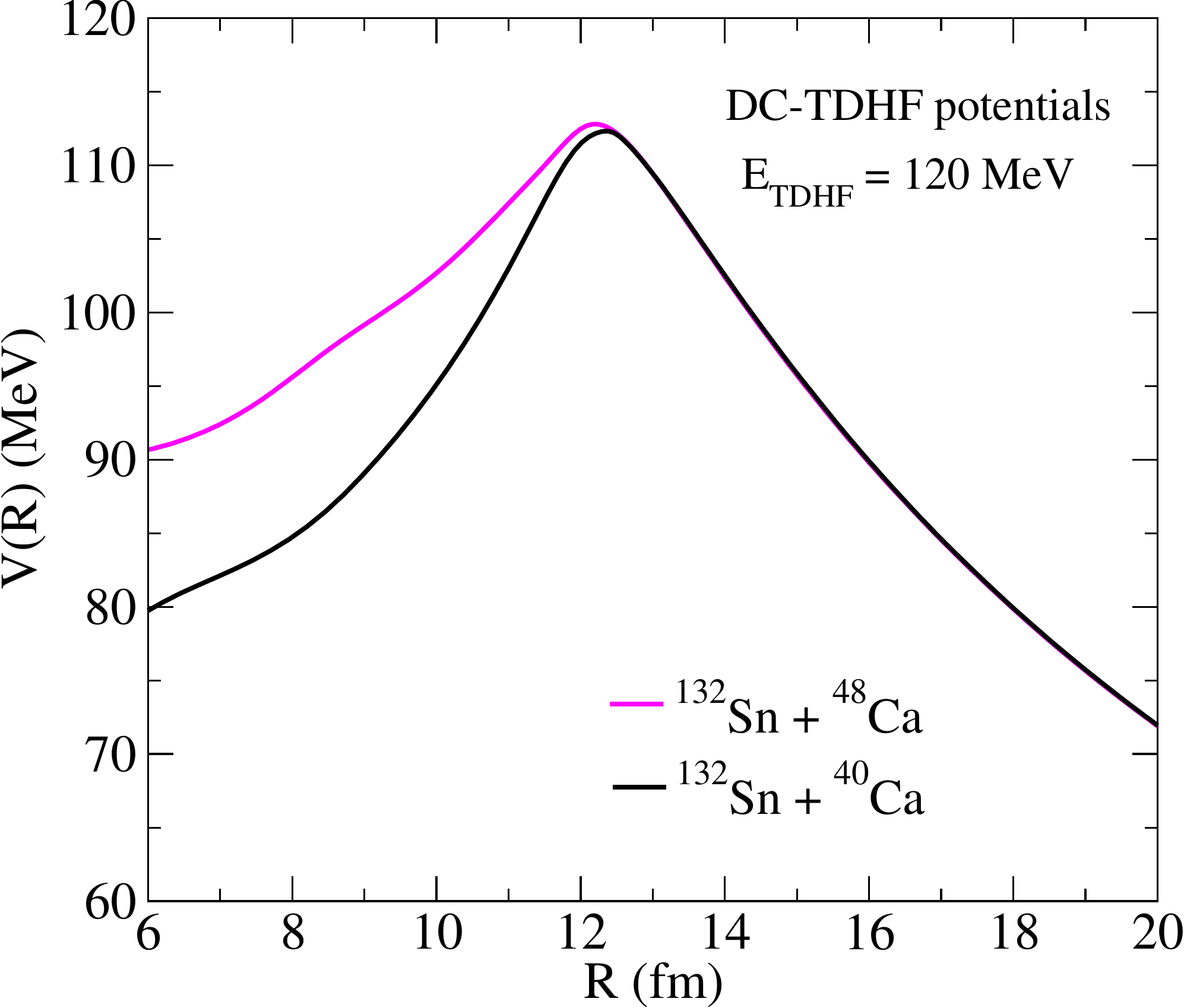}\hspace{0.5cm}
\includegraphics[width=6.0cm]{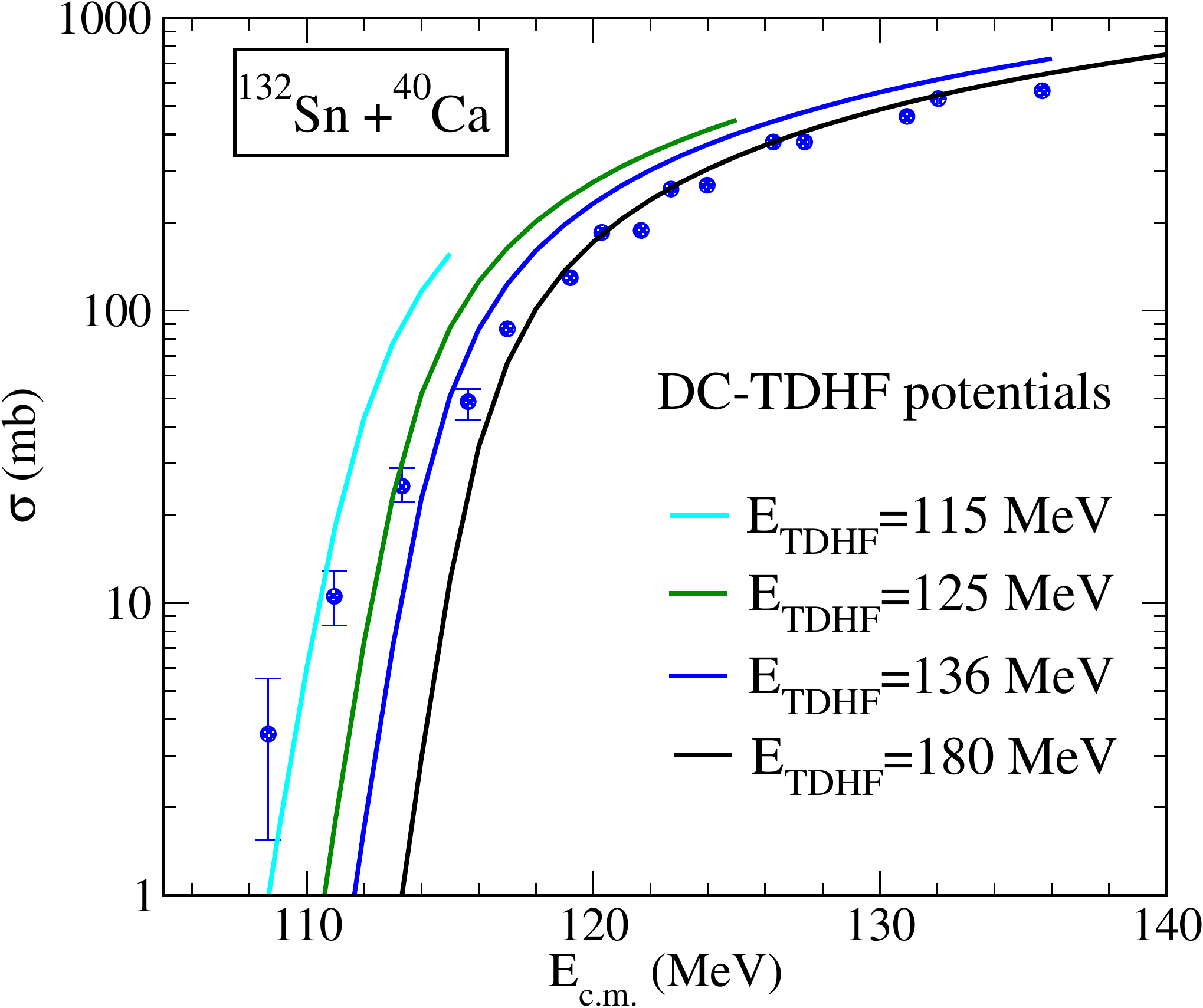}
\end{center}
\caption{\protect\footnotesize Left: Comparison of DC-TDHF heavy-ion interaction
potentials, calculated at $E_\mathrm{c.m.}=120$ MeV, for the systems $^{132}$Sn+$^{40}$Ca
and $^{132}$Sn+$^{48}$Ca. Right: Total fusion cross sections for $^{132}$Sn+$^{40}$Ca
as a function of center-of-mass energy. The cross sections have been calculated for
several energy-dependent potentials. The experimental data are taken
from Ref.~\cite{kolata2012}.}
\label{fig:Sn+Ca_2}
\end{figure}
Using the microscopic DC-TDHF approach
we are able to explain the measured sub-barrier fusion enhancement~\cite{oberacker2013} in terms
of the {\it narrower width} of the ion-ion potential for $^{132}$Sn+$^{40}$Ca,
while the barrier heights and positions are approximately the same in both systems,
see Fig.~\ref{fig:Sn+Ca_2}. The energy-dependence of the heavy-ion interaction
potentials is crucial for understanding the low-energy fusion data.

\begin{figure}[!htb]
\begin{center}
\includegraphics[width=6.0cm]{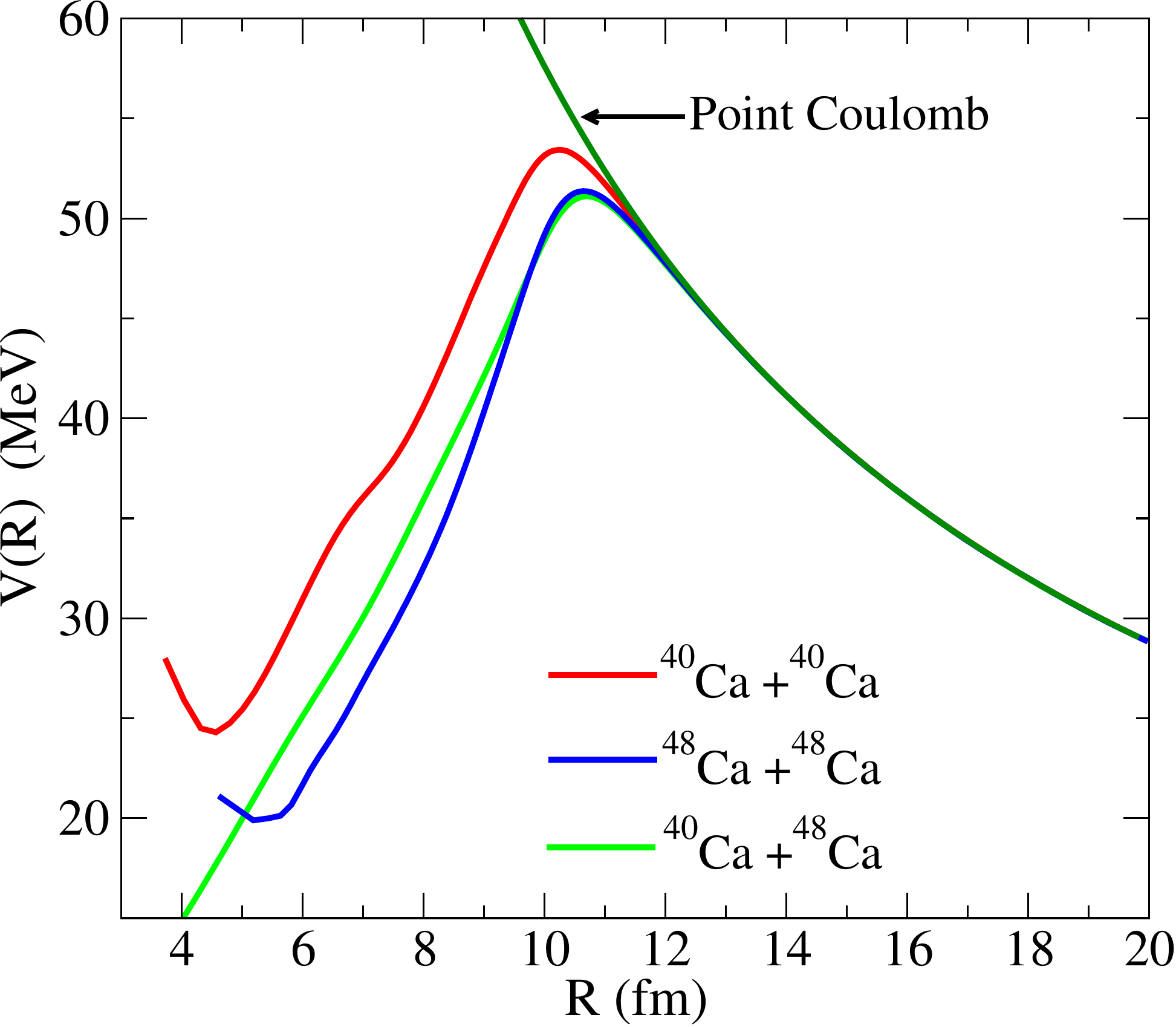}\hspace{0.5cm}
\includegraphics[width=6.0cm]{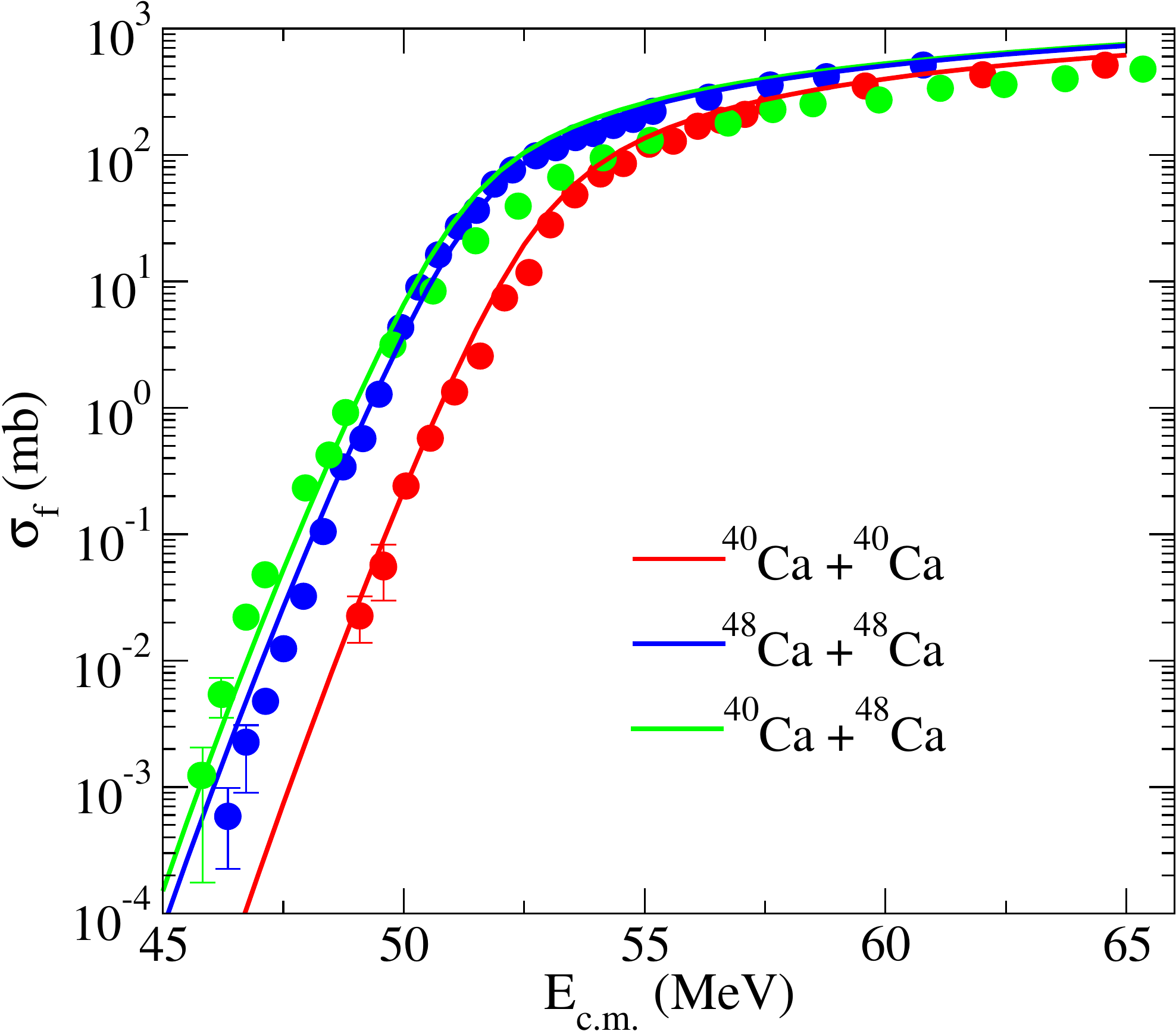}
\end{center}
\caption{\protect\footnotesize DC-TDHF ion-ion potentials (left) and fusion cross-sections (right)
for various isotopes of the Ca$+$Ca system~\cite{keser2012}. The experimental data are taken from Ref.~\cite{montagnoli2012}.}
\label{fig:Ca+Ca}
\end{figure}

As another example, we discuss our theoretical analysis of recent
$^{40,48}$Ca +$^{40,48}$Ca fusion data~\cite{montagnoli2012} which extend older fusion data
to low sub-barrier energies. Comparison of the sub-barrier
cross sections with those calculated using standard coupled-channel calculations
suggested a hindrance of the fusion cross-sections at deep sub-barrier energies.
The DC-TDHF results for the ion-ion potentials and fusion cross section~\cite{keser2012} are shown in Fig.~\ref{fig:Ca+Ca}.
In contrast to the heavy systems discussed above, we find that the ion-ion potentials
for lighter systems depend only weakly on the energy, and the potential barrier
corresponding to the lowest collision energy gives the best fit to the sub-barrier cross-sections
since this is the one that allows for more rearrangements to take place and grows the inner part
of the barrier. Considering the fact that historically the low-energy sub-barrier cross-sections
of the $^{40}$Ca+$^{48}$Ca system have been the ones not reproduced well by the standard models,
the DC-TDHF results are quite satisfactory, indicating that the dynamical evolution of the
nuclear density in TDHF gives a good overall description of the collision process.

\begin{figure}[!htb]
\begin{center}
\includegraphics[width=8.0cm]{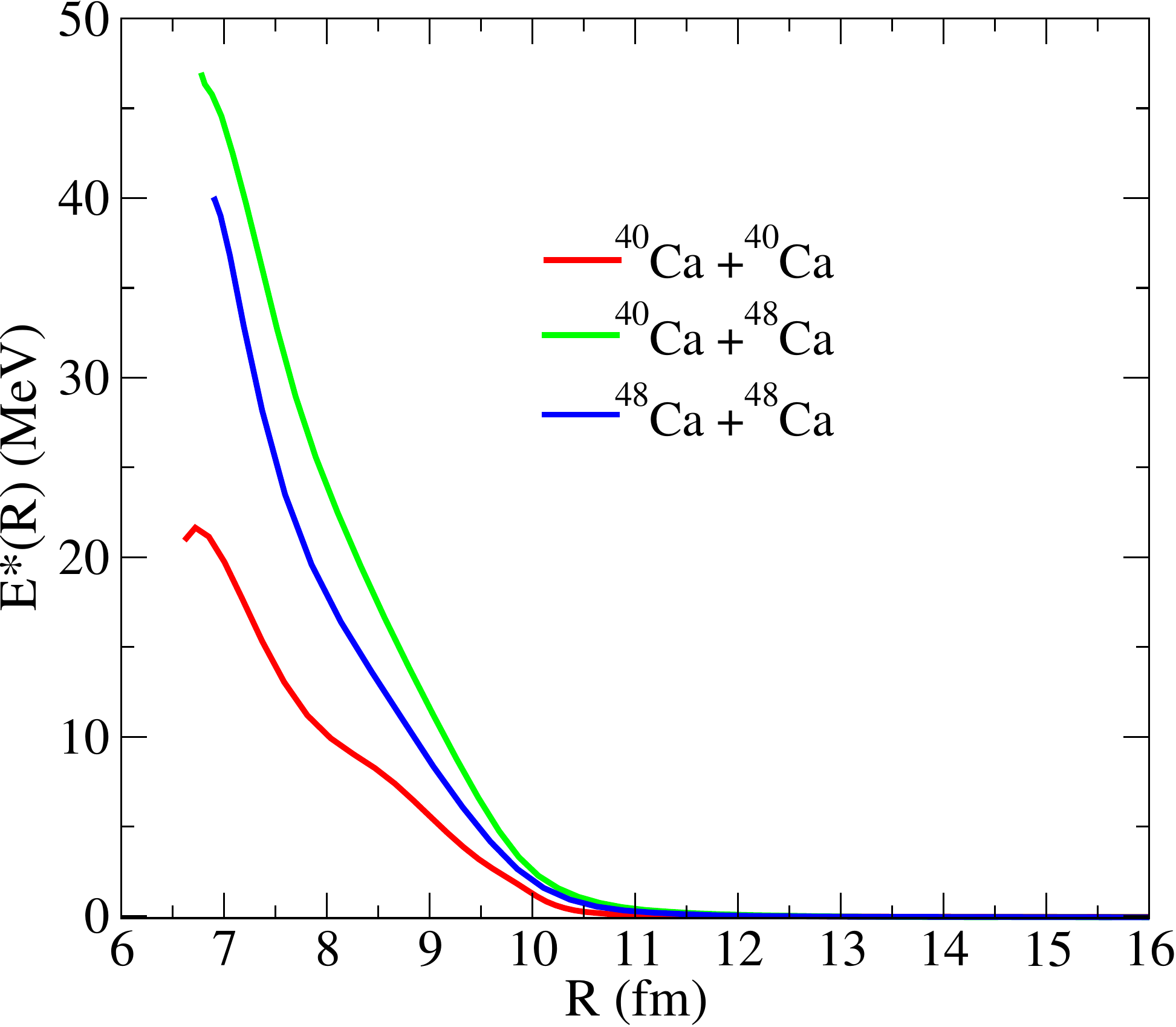}
\end{center}
\caption{\protect\footnotesize Excitation energy $E^*(R)$, for three isotopic
combinations of the Ca$+$Ca system~\cite{keser2012}.}
\label{fig:estar_Ca+Ca}
\end{figure}

Figure~\ref{fig:estar_Ca+Ca} shows the excitation energy, $E^*(R)$, for the three systems
studies here. The excitation energy was calculated for the same value of
$\varepsilon=E_\mathrm{c.m.}/\mu=2.75$~MeV
for all systems, which corresponds to collision energies of $55$, $60$, and $66$~MeV, respectively.
All curves initially behave in a similar manner, at large distances the excitation is zero, as the
nuclei approach the barrier peak the excitations start and monotonically rise for larger overlaps.
The interesting observation is that the excitations for the intermediary $^{40}$Ca+$^{48}$Ca system
start at a slightly earlier time and rise above the other two systems.
This may be largely due to the fact that an asymmetric system has some additional modes
of excitation in comparison to the other two symmetric systems.

%---------------------------------------------------------------

\subsection{Hot and Cold Fusion to Produce Superheavy Elements}

\begin{figure}[!htb]
\begin{center}
\includegraphics[width=6.0cm]{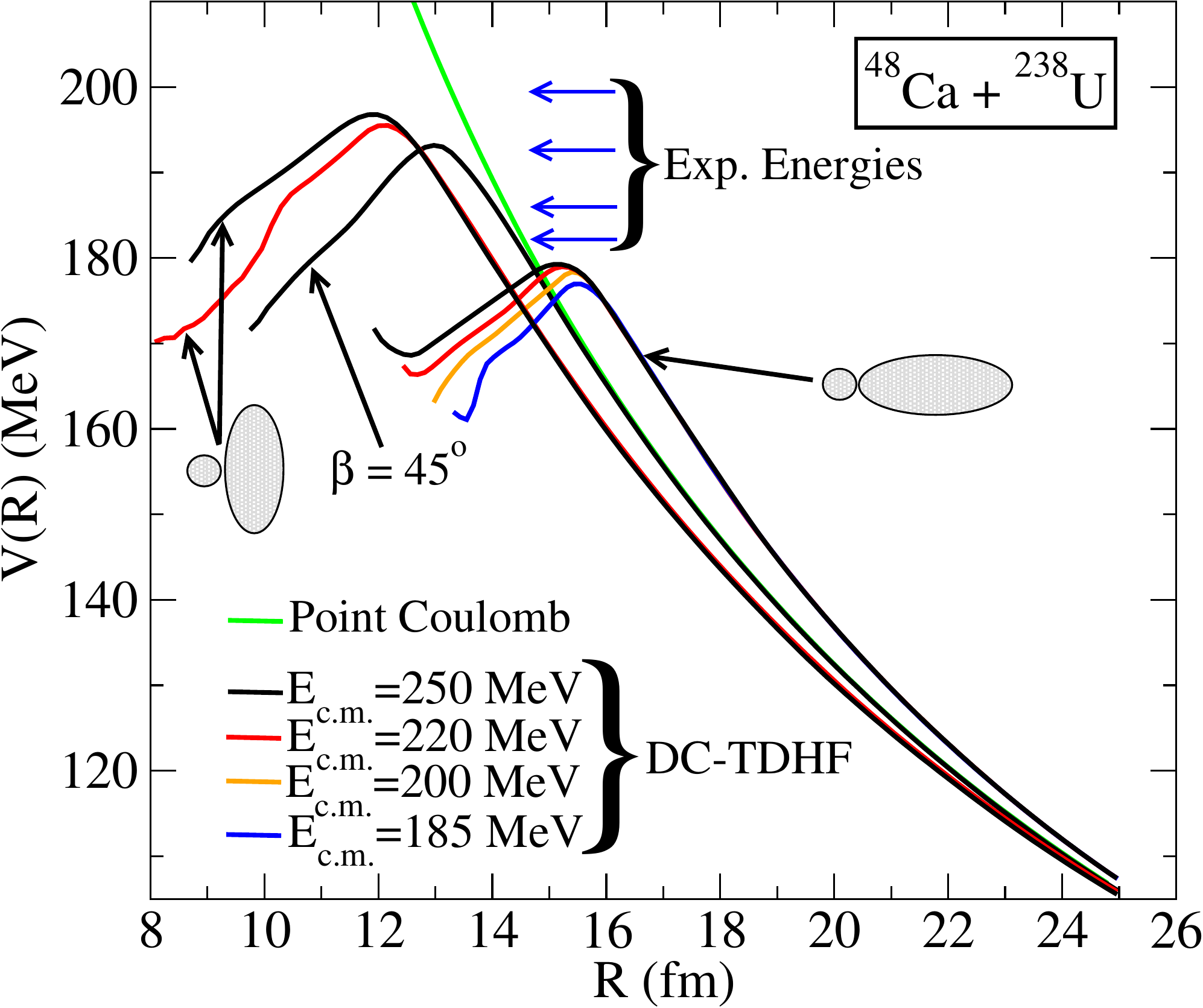}\hspace{0.5cm}
\includegraphics[width=6.0cm]{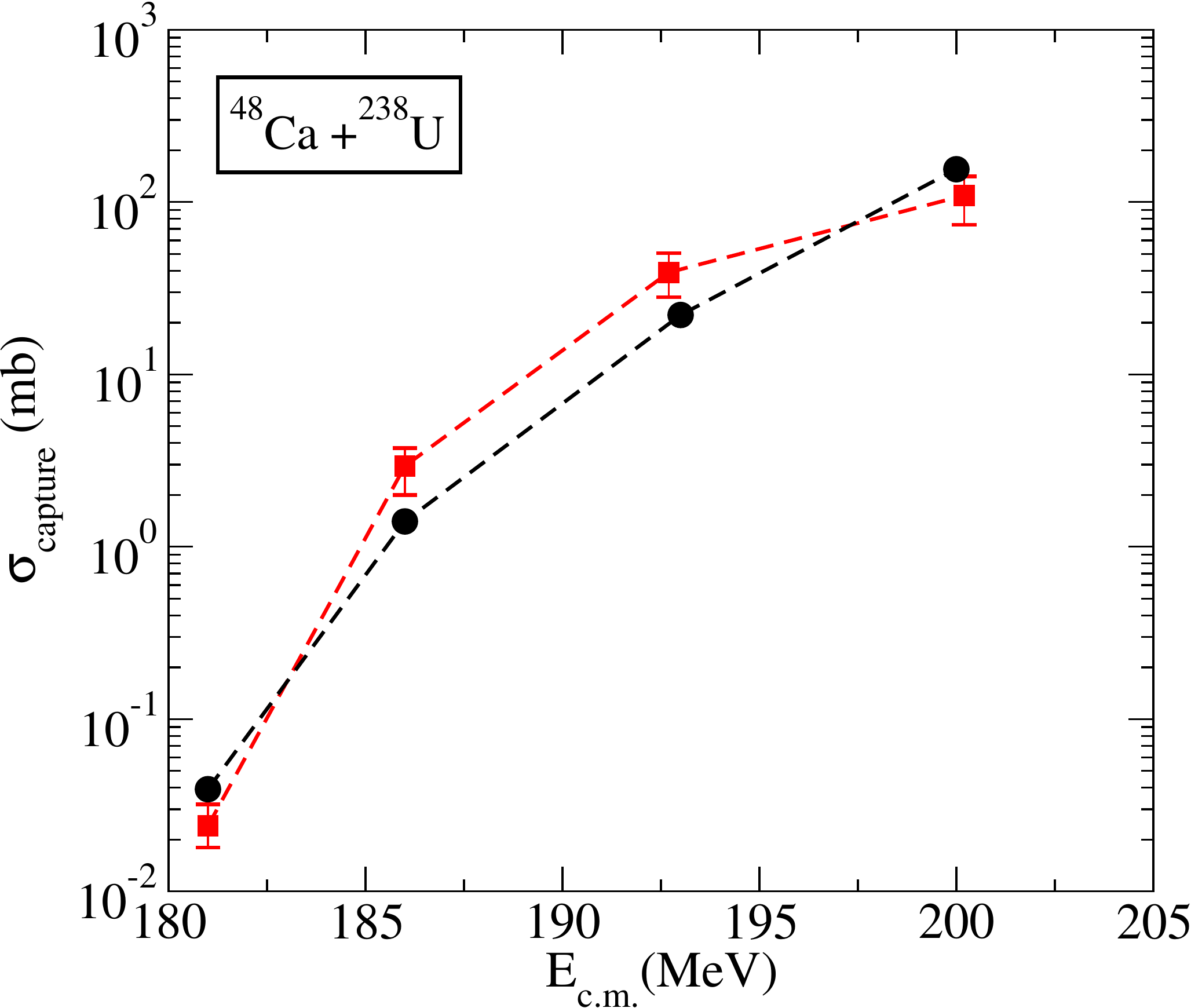}
\end{center}
\caption{\protect\footnotesize Left: Potential barriers, $V(R)$, obtained from
DC-TDHF calculations~\cite{umar2010a} as a function of $E_\mathrm{c.m.}$ energy and
orientation angle $\beta$ of the $^{238}$U nucleus. Also shown are the experimental c.m. energies.
Right: Capture cross-sections as a function of $E_\mathrm{c.m.}$ energy (black circles).
Also shown are the experimental cross-sections~\cite{itkis2007,oganessian2007} (red squares).}
\label{fig:vrCa_U}
\end{figure}

In connection with superheavy element production, we have studied the hot fusion reaction
$^{48}$Ca+$^{238}$U and the cold fusion reaction $^{70}$Zn+$^{208}$Pb~\cite{umar2010a}.
Considering hot fusion,
$^{48}$Ca is a spherical nucleus whereas $^{238}$U has a large axial deformation.
The deformation of $^{238}$U strongly influences the interaction barrier for this
system. This is shown in Fig.~\ref{fig:vrCa_U}, which shows the interaction barriers, $V(R)$,
calculated using the DC-TDHF method as a function of c.m. energy and for three different
orientations of the $^{238}$U nucleus. The alignment angle $\beta$ is the angle
between the symmetry axis of the $^{238}$U nucleus and the collision axis.

The barriers for the polar orientation ($\beta=0^{o}$) of the
$^{238}$U nucleus are much lower and peak at larger ion-ion separation distance $R$.
On the other hand, the barriers for the equatorial orientation ($\beta=90^{o}$) are
higher and peak at smaller $R$ values.
We observe that at lower energies the polar orientation results in
sticking of the two nuclei, while the equatorial orientation results in a deep-inelastic
collision.
We have also calculated the excitation energy $E^{*}(R)$ as a function of c.m. energy
and orientation angle $\beta$ of the $^{238}$U nucleus.
The system is
excited much earlier during the collision process for the polar orientation
and has a higher excitation than the corresponding collision for the equatorial orientation.

To obtain the capture cross-section, we calculate potential barriers $V(R,\beta)$
for a set of initial orientations $\beta$ of the $^{238}$U nucleus. Then we determine
partial cross sections $\sigma(\beta)$ and perform an angle-average, including
the dynamic alignment arising from Coulomb excitation of the g.s. rotational band
in the entrance channel. In Fig.~\ref{fig:vrCa_U} we show our
results for the capture cross-sections which are in remarkably good agreement
with experimental data.

One of the major questions that is asked by the experimental superheavy element community is
why a $^{48}$Ca beam is so crucial in forming such systems and whether
one could produce new superheavy nuclei using projectiles different than
$^{48}$Ca and actinide targets.
Some possible projectiles include $^{50}$Ti, $^{54}$Cr, $^{56}$Fe and $^{64}$Ni.
Several reactions with these projectiles were tried, no SHE event reported so far, only cross section limits.
Typical reactions studied already are $^{238}$U +$^{64}$Ni (SHIP GSI), $^{248}$Cm+$^{54}$Cr (SHIP)
and $^{248}$Cm+$^{56}$Fe (Dubna), $^{50}$Ti+$^{249}$Bk and $^{50}$Ti+$^{249}$Cf (TASCA GSI).
During the ICFN5 meeting at Sanibel Island, Florida a particularly
low cross section limit,
of about $50$~fb, was reported for the production of isotopes of new element $119$ in the reaction between
$^{50}$Ti and $^{249}$Bk. The reaction $^{48}$Ca +$^{249}$Bk makes superheavy isotopes of
element $117$ with cross-sections of $2-3$ picobarns, by contrast the $^{50}$Ti +$^{249}$Bk
reaction yields a cross-section limit of $50$~fb, about $50$ times lower.
TDHF can be used to calculate
capture cross-sections, excitation energies, and other dynamical
properties for some of these systems in order to find the
discerning physical property among them.

%---------------------------------------------------------------

\subsection{Nuclear Astrophysics: Fusion of Oxygen and Carbon in the Neutron Star Crust}

\begin{figure}[!htb]
\begin{center}
\includegraphics[width=6.0cm]{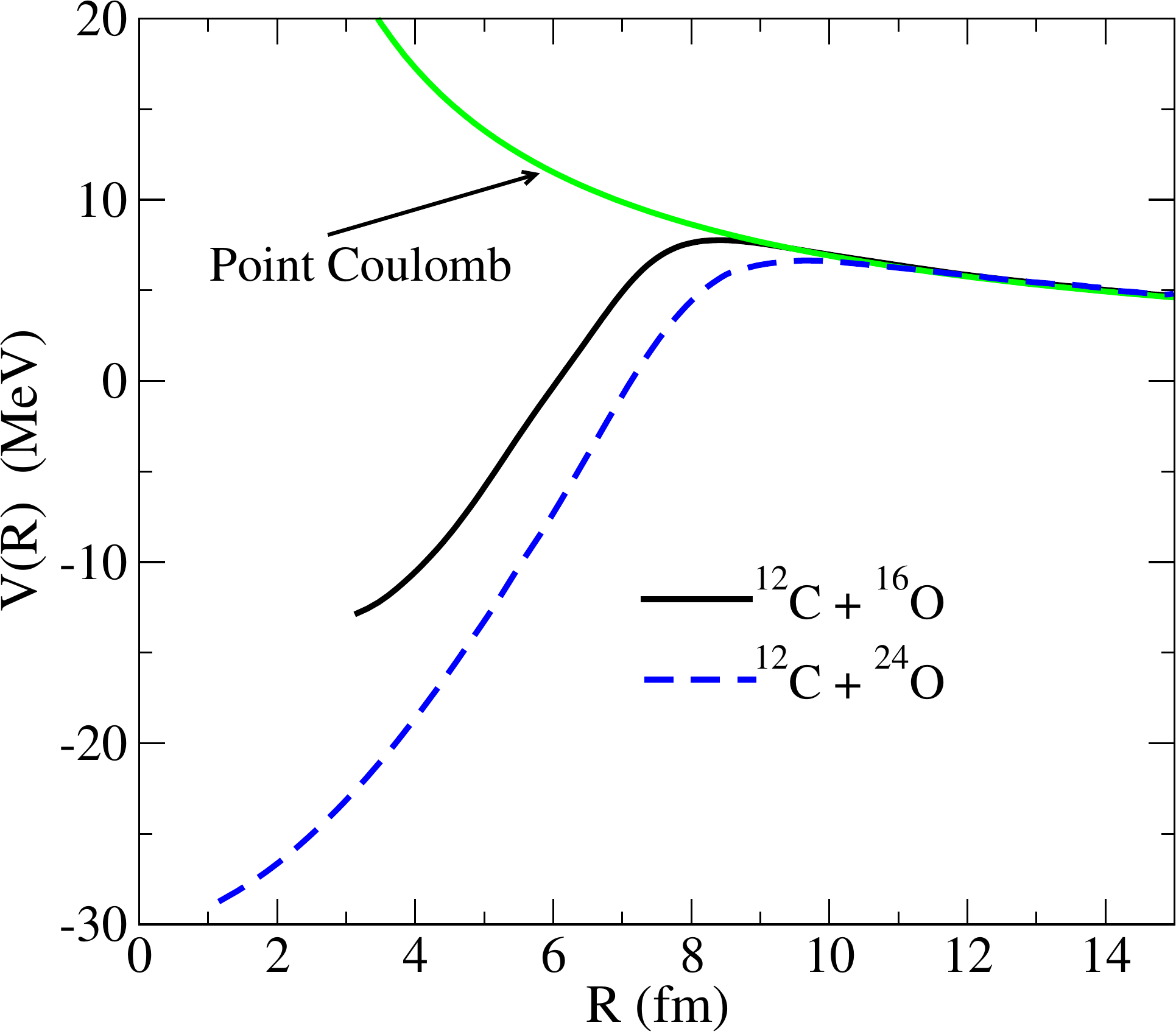}\hspace{0.5cm}
\includegraphics[width=6.0cm]{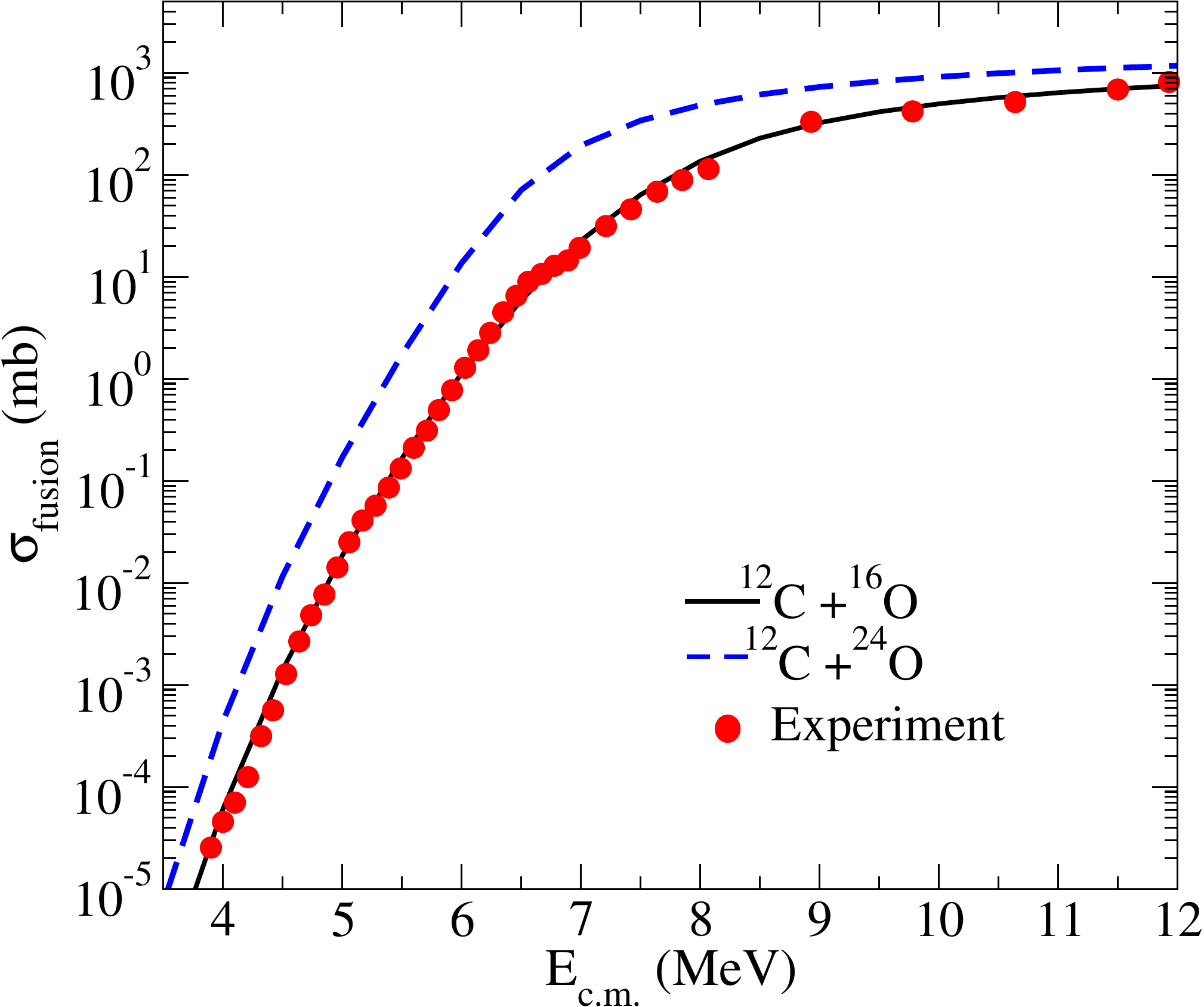}
\end{center}
\caption{\protect\footnotesize Left: DC-TDHF heavy-ion potentials for the systems $^{12}$C+$^{16,24}$O.
Right: Total fusion cross sections versus center-of-mass
energy for fusion of carbon with oxygen isotopes~\cite{umar2012a}. The experimental data are from Ref.~\cite{jiang2007}.}
\label{fig:C+O}
\end{figure}

Fusion of very neutron rich nuclei may be important to determine the composition and
heating of the crust of accreting neutron stars.
We have studied sub-barrier fusion reactions between both stable and neutron-rich
isotopes of oxygen and carbon. In Fig.~\ref{fig:C+O} (left side) we show the DC-TDHF
potential barriers for the C$+$O system~\cite{umar2012a}.
The higher barrier corresponds to the $^{12}$C$+$ $^{16}$O system and has a peak
energy of $7.77$~MeV. The barrier for the $^{12}$C$+$ $^{24}$O system occurs at a
slightly larger $R$ value with a barrier peak of $6.64$~MeV. The right side of
Fig.~\ref{fig:C+O} shows the corresponding cross sections for the two reactions.
Also shown are the experimental data from Ref.~\cite{jiang2007}. The
DC-TDHF potential reproduces the experimental cross-sections quite well for the
$^{12}$C$+$ $^{16}$O system, and the cross section for the neutron rich $^{12}$C+$^{24}$O
is predicted to be larger than that for $^{12}$C +$^{16}$O. Similar agreement with
fusion data for the $^{16}$O$+$ $^{16}$O system and the predicted cross-sections
for heavier oxygen isotopes were also calculated~\cite{umar2012a}.

%---------------------------------------------------------------

\section*{Summary}

In this manuscript we have outlined the microscopic study of fusion using
the DC-TDHF approach.
For a variety of heavy-ion reactions,
we discuss the results of our numerical calculations of fusion / capture cross-sections
at energies below and above the Coulomb barrier and compare these to
experimental data if available. In this context we discuss a variety of stable and
neutron-rich reaction partners, and we examine hot fusion
reactions leading to the formation of superheavy elements.
Fusion of very neutron rich light nuclei also plays a major role in
nuclear astrophysics where it determines the composition and heating of the crust of
accreting neutron stars.
We have shown that microscopically obtained ion-ion potentials do give a
reasonably good description of these fusion cross-sections.

The fully microscopic TDHF theory has shown itself to be rich in
nuclear phenomena and continues to stimulate our understanding of nuclear dynamics.
The time-dependent mean-field studies seem to show that the dynamic evolution
builds up correlations that are not present in the static theory.
While modern Skyrme forces provide a much better description of static nuclear properties
in comparison to the earlier parametrizations there is a need to obtain even better
parametrizations that incorporate deformation and reaction data into the fit process.

%---------------------------------------------------------------

\section*{Acknowledgment}
We would like to extend our best wishes to Prof. Joachim Maruhn for his retirement.
Since the $1970$'s Prof. Maruhn has made significant contributions to the development
of time-dependent mean-field theories and lead the way for the new generation of nuclear
physicists to build upon this foundation. Both as a colleague and a friend we would like
to extend our gratitude to him.
This work has been supported by the U.S. Department of Energy under grant No.
DE-FG02-96ER40975 with Vanderbilt University.

\bibliography{VU_bibtex_master.bib}

\end{document}